\documentclass[twocolumn,aps,showkeys,amsmath,amssymb]{revtex4}
\usepackage{graphicx}
\usepackage{dcolumn}
\usepackage{amsmath}%
\usepackage{amssymb}%

\begin{document}


\title{Derivation of generalized Einstein's equations of gravitation in inertial systems based on a sink flow model of particles}

\author{Xiao-Song Wang}
\affiliation{Institute of Mechanical and Power Engineering, Henan Polytechnic University,
Jiaozuo, Henan Province, 454000, China}
\date{Jul. 19, 2019}

\begin{abstract}
J. C. Maxwell, B. Riemann and H. Poincar$\acute{e}$ have proposed the idea that all microscopic particles are sink flows in a fluidic aether. Following this research program, a previous theory of gravitation based on a mechanical model of vacuum and a sink flow model of particles is generalized by methods of special relativistic continuum mechanics. In inertial coordinate systems, we construct a tensorial potential which satisfies the wave equations. Inspired by the equations of motion of a test particle, a definition of a metric tensor of a Riemannian spacetime is introduced. Applying Fock's theorem, generalized Einstein's equations in inertial systems are derived based on some assumptions. These equations reduce to Einstein's equations in case of weak field in harmonic coordinate systems.
\end{abstract}


\maketitle


\section{Introduction  \label{sec 100}}
\newtheorem{assumption_my}{\bfseries Assumption}

\newtheorem{definition_my}{\bfseries Definition}

\newtheorem{lemma_my}{\bfseries Lemma}

\newtheorem{proposition_my}{\bfseries Proposition}

\newtheorem{theorem_my}{\bfseries Theorem}

\newtheorem{wcorollary_my}{\bfseries Corollary}

The Einstein's equations of gravitational fields in the theory of general relativity can be written as \cite{WeinbergS1972,MisnerC1973}
\begin{equation}\label{Einstein 100-100}
R_{\mu\nu}-\frac{1}{2}g_{\mu\nu}R=-\kappa T^{m}_{\mu\nu},
\end{equation}
where $g_{\mu\nu}$ is the metric tensor of a Riemannian spacetime,
$R_{\mu\nu}$ is the Ricci tensor, $R\equiv g^{\mu\nu}R_{\mu\nu}$ is the scalar curvature, $g^{\mu\nu}$ is the contravariant metric tensor, $\kappa$ is a constant, $T^{m}_{\mu\nu}$ is the energy-momentum tensor of a matter system.

The Einstein's equations (\ref{Einstein 100-100}) are fundamental assumptions in the theory of general relativity \cite{WeinbergS1972,MisnerC1973}. It is remarkable that Einstein's theory of general relativity, born in the year of 1915, has held up under every experimental test, refers to, for instance, \cite{HeesA20170526}.

There is a long history of researches of derivations or interpretations of Einstein's theory of general relativity.
For instance, C. Misner et al. introduce six derivations of the Einstein's equations Eqs.\ (\ref{Einstein 100-100}) in their great book (\cite{MisnerC1973}, p.\ 417). S. Weinberg proposed two derivations (\cite{WeinbergS1972}, p.\ 151).

However, these theories still face the following difficulties. (1) Attempts to reconcile the theory of general relativity and quantum mechanics have met some mathematical difficulties (\cite{MaddoxJ1998},p.\ 101); (2) The cosmological constant problem is still a puzzle, refers to, for instance, \cite{MarshD2017}; (3) The existence of black hole is still controversy, refers to, for instance, \cite{ChoA2017}; (4) Theoretical interpretation of P. A. M. Dirac's dimensionless large number (\cite{DiracP1978}, p.\ 73) is still open; (5) The existences and characters of dark matter and dark energy are still controversy, refers to, for instance,  \cite{KimJ2017}; (6) The existence and characters of gravitational aether are still not clear, refers to, for instance, \cite{FarhoudiM2016}; (7) Whether Newton's gravitational constant $\gamma_{N}$ depends on time and space is still not clear \cite{LongDR1976}; (8) Whether the speed of light in vacuum depends on time or space is controversy, refers to, for instance,  \cite{BalcerzakA2017}.

Furthermore, there exists some other problems related to the theories of gravity, for instance, the definition of inertial system, origin of inertial force, the velocity of the propagation of gravity \cite{CornishN2017}, the velocity of individual photons \cite{GiovanniniD2015,SamblesJR2015}, unified field theory, etc.

The purpose of this manuscript is to propose a derivation of the Einstein's equations (\ref{Einstein 100-100}) in inertial coordinate systems based on a mechanical model of vacuum and a sink flow model of particles \cite{WangXS200810}.

\section{Introduction of a previous theory of gravitation based on a sink flow model of particles by methods of classical fluid mechanics \label{sec 200}}
The idea that all microscopic particles are sink flows in a fluidic substratum is not new. For instance, in order to compare fluid motions with electric fields, J. C. Maxwell introduced an analogy between source or sink flows and electric charges (\cite{WhittakerE1951}, p.\ 243). B. Riemann speculates that:"{\itshape  I make the hypothesis that space is filled with a substance which continually flows into ponderable atoms, and vanishes there from the world of phenomena, the corporeal world}"(\cite{RiemannB2004}, p.\ 507). H. Poincar$\acute{e}$ also suggests that matters may be holes in fluidic aether (\cite{PoincareH1997}, p.\ 171). A. Einstein and L. Infeld said (\cite{EinsteinAInfeldL}, p.\ 256-257):"{\itshape Matter is where the concentration of energy is great, field where the concentration of energy is small. $\cdots$ What impresses our senses as matter is really a great concentration of energy into a comparatively small space. We could regard matter as the regions in space where the field is extremely strong.}"

Following these researchers, we suppose that all the microscopic particles were made up of a kind of elementary sinks of a fluidic medium filling the space \cite{WangXS200810}. Thus, Newton's law of gravitation is derived by methods of hydrodynamics based on the fluid model of vacuum and the sink flow model of particles \cite{WangXS200810}.

We briefly introduce this theory of gravitation \cite{WangXS200810}.
Suppose that there exists a fluidic medium filling the interplanetary vacuum.
For convenience, we may call this medium as the $\Omega(0)$ substratum, or gravitational aether, or tao \cite{WangXS200810}.
Suppose that the following conditions are valid: (1) the $\Omega(0)$ substratum is
an ideal fluid;  (2) the ideal fluid is irrotational and barotropic;
(3) the density of the $\Omega(0)$ substratum is homogeneous;
(4) there are no external body forces exerted on the fluid;
(5) the fluid is unbounded and the velocity of the fluid at the infinity is approaching to zero.

\begin{figure}
\begin{center}
\resizebox*{4cm}{!}{\includegraphics{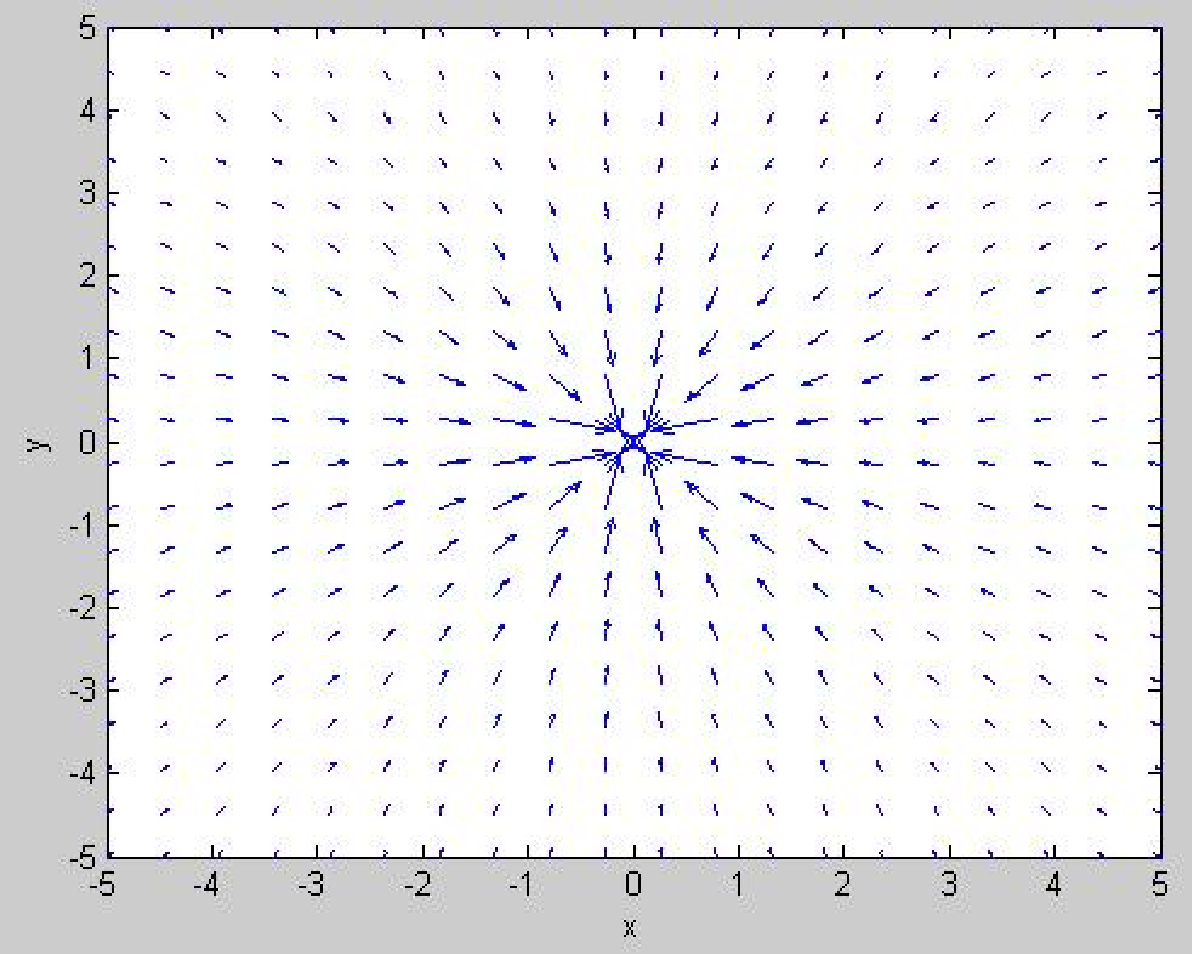}}
\caption{an illustration of the velocity field of a sink flow.} \label{sink 200-100}
\end{center}
\end{figure}

An illustration of the velocity field of a sink flow can be found in Figure \ref{sink 200-100}.
If a point source is moving with a velocity $\mbox{\upshape\bfseries{v}}_s$, then there is a force \cite{WangXS200810}
\begin{equation}\label{force 200-100}
\mbox{\upshape\bfseries{F}}_Q= -\rho_{0}Q(\mbox{\upshape\bfseries{u}}-\mbox{\upshape\bfseries{v}}_s)
\end{equation}
is exerted on the source by the fluid, where $\rho_{0}$ is the density
of the fluid, $Q$ is the strength of the source,
$\mbox{\upshape\bfseries{u}}$ is the velocity of the fluid at the
location of the source induced by all means other than the source
itself.

We suppose that all the elementary sinks were created simultaneously \cite{WangXS200810}. For convenience, we may call these elementary sinks as monads. The initial masses and the strengths of the monads are the same. Suppose that
(1) $\mbox{\upshape\bfseries{v}}_{i}\ll\mbox{\upshape\bfseries{u}}_{i}, i=1, 2$, where
$\mbox{\upshape\bfseries{v}}_{i}$ is the velocity of the particle with
mass $m_{i}$, $\mbox{\upshape\bfseries{u}}_i$ is the velocity of the
$\Omega(0)$ substratum at the location of the particle with mass
$m_i$ induced by the other particle;
 (2) there are no other forces exerted on the particles,
then the force $\mbox{\upshape\bfseries{F}}_{12}(t)$
 exerted on the particle with mass $m_2(t)$
 by the velocity field of $\Omega(0)$ substratum induced by the particle with mass $m_1(t)$
 is \cite{WangXS200810}
\begin{equation}\label{gravitation 200-200}
\mbox{\upshape\bfseries{F}}_{12}(t)=- \gamma_{N}(t)\frac{m_1(t)
m_2(t)}{r^2}\hat{\mbox{\upshape\bfseries{r}}}_{12},
\end{equation}
where$\gamma_{N}$ is Newton's gravitational constant, $\hat{\mbox{\upshape\bfseries{r}}}_{12}$ denotes the unit vector
directed outward along the line from the particle with mass $m_1(t)$
to the particle with mass $m_2(t)$, $r$ is the distance between the
two particles, $m_0(t)$ is the mass of monad at time $t$, $-q_0( q_0 > 0)$ is
the strength of a monad, and
\begin{equation}\label{constant 200-210}
\gamma_{N}(t)=\frac{\rho_{0} q^2_0}{4\pi m^2_0(t)}.
\end{equation}

For continuously distributed matter, we have
\begin{equation}\label{conservation 200-300}
 \frac{\partial \rho_{0}}{\partial t}
 + \nabla \cdot (\rho_{0} \mbox{\bfseries{u}}) = -\rho_{0}\rho_{s},
\end{equation}
where
$\mbox{\bfseries{u}}$ is the velocity of the $\Omega(0)$ substratum, $\nabla =
\mbox{\upshape\bfseries{i}}\partial /\partial x +
\mbox{\upshape\bfseries{j}}\partial /\partial y +
\mbox{\upshape\bfseries{k}}\partial /\partial z$ is the
nabla operator introduced by Hamilton, $\mbox{\upshape\bfseries{i}}, \mbox{\upshape\bfseries{j}},
\mbox{\upshape\bfseries{k}}$ are basis vectors,
$-\rho_{s} ( \rho_{s}>0 )$ is the density of continuously distributed sinks, i.e.,
\begin{equation}\label{density 600-300}
-\rho_{s}=\lim_{\triangle V \rightarrow 0}\frac{\triangle Q}{\triangle V},
\end{equation}
where $\triangle Q$ is the source strength of the continuously distributed matter
in the volume $\triangle V$ of the $\Omega(0)$ substratum.

Since the $\Omega(0)$ substratum is homogeneous, i.e.,
$\partial\rho_{0}/\partial t=\partial\rho_{0}/\partial x
=\partial\rho_{0}/\partial y=\partial\rho_{0}/\partial z
=0$, and irrotational, i.e., $\nabla \times \mbox{\bfseries{u}}=0$,
Eq.\ (\ref{conservation 200-300}) can be written as \cite{Currie2003}
\begin{equation}\label{conservation 200-400}
\nabla^{2}\varphi= -\rho_{s},
\end{equation}
where $\varphi$ is a velocity potential such that $\mbox{\bfseries{u}}=\nabla \varphi$, $\nabla^2 = \partial^2
/\partial x^2 + \partial^2 /\partial y^2 +
\partial^2 /\partial z^2$ is the Laplace operator.

We introduce the following definitions
\begin{equation}\label{definition 200-500}
\Phi=\frac{\rho_{0} q_0}{m_0}\varphi, \quad  \rho_m =  \frac{m_0 \rho_s}{q_0},
\end{equation}
where $\rho_{m}$ denotes the mass density of continuously distributed particles.

Using Eq.\ (\ref{definition 200-500}) and Eq.\ (\ref{constant 200-210}),
Eq.\ (\ref{conservation 200-400}) can be written as
\begin{equation}\label{conservation 200-600}
\nabla^{2}\Phi = -4\pi \gamma_{N} \rho_{m}.
\end{equation}

\section{A mechanical model of vacuum \label{sec 300}}
According to our previous paper \cite{WangXS200804} we suppose that vacuum is filled with a kind of continuously distributed material which may be called $\Omega(1)$ substratum or electromagnetic aether.
Maxwell's equations in vacuum are derived by methods of
continuum mechanics based on this mechanical model of vacuum and a source and sink flow model of electric charges \cite{WangXS200804}. We speculate that the electromagnetic aether may also generate gravity. Thus, we introduce the following assumption.
\begin{assumption_my}\label{assumption 300-200}
The particles that constitute the $\Omega(1)$ substratum, or the electromagnetic aether,
are sinks in the $\Omega(0)$ substratum.
\end{assumption_my}

Then, according to the previous theory of gravitation \cite{WangXS200810},
these $\Omega(1)$ particles gravitate with each other and also attract with matters.
Thus, vacuum is composed of at least two kinds of interacting substratums, i.e.,
the gravitational aether $\Omega(0)$ and the electromagnetic aether $\Omega(1)$.

From Eq.\ (\ref{force 200-100}), we see that there exists a following universal damping force $\mathbf{F}_d= -\rho_{0} q_0m\mathbf{v}_{p}/m_0$ exerted on each particle by the $\Omega(0)$ substratum \cite{WangXS200810},
where $\mathbf{v}_{p}$ is the velocity of the particle.
Based on this universal damping force $\mathbf{F}_d$ and some assumptions, we derive a generalized Schr\"{o}dinger equation for microscopic particles \cite{WangXS2014PhysicsEssays}. For convenience, we may call these theories \cite{WangXS200804,WangXS200810,WangXS2014PhysicsEssays} as the theory of vacuum mechanics.

\section{Construction of a Lagrangian for free fields of the $\Omega(0)$ substratum based on a tensorial potential in the Galilean coordinates \label{sec 400}}
There exists some approaches (\cite{FeynmanRP1995}, page vii;\cite{MisnerC1973}, p.\ 424), which regards Einstein's general relativity as a special relativistic field theory in an unobservable flat spacetime, to derive the Einstein's equations (\ref{Einstein 100-100}). However, these theories can not provide a physical definition of the tensorial potential of gravitational fields, refers to, for instance, \cite{ThirringW1961,WeinbergS1965,MisnerC1973}. Thus, similar to the theory of general relativity, these theories may be regarded as phenomenological theories of gravitation.

Inspired by these special relativistic field theories of gravitation, we explore the possibility of establishing a similar theory based on the theory of vacuum mechanics \cite{WangXS200804,WangXS200810,WangXS2014PhysicsEssays}. Thus, first of all, we need to construct a Lagrangian for free fields of the $\Omega(0)$ substratum based on a tensorial potential in the Galilean coordinates. In this section, we will regard the $\Omega(0)$ substratum in the previous theory of gravitation \cite{WangXS200810} as a special relativistic fluid. Then, we will study the $\Omega(0)$ substratum by methods of special relativistic continuum mechanics \cite{MollerC1955}.

In this article, we adopt the mathematical framework of the theory of special relativity \cite{WeinbergS1972}. However, the physical interpretation of the mathematics of the theory of special relativity may be different from Einstein's theory. It is known that Maxwell's equations are valid in the frames of reference that attached to the $\Omega(1)$ substratum \cite{WangXS200804}. We introduce a Cartesian coordinate system $\{ 0, x, y, z \}$ for a three-dimensional Euclidean space that attached to the $\Omega(1)$ substratum. Let $\{ 0, t \}$ be a one-dimensional time coordinate. We denote this reference frame as $S_{\Omega(1)}$.

Based on the Maxwell's equations, the law of propagation of an electromagnetic wave front
in this reference frame $S_{\Omega(1)}$ can be derived and can be written as (\cite{FockV1964},p.\ 13)
\begin{equation}\label{front 400-400}
\frac{1}{c^{2}}\left ( \frac{\partial \omega}{\partial t} \right )^{2}-
\left ( \frac{\partial \omega}{\partial x} \right )^{2}
-\left ( \frac{\partial \omega}{\partial y} \right )^{2}
-\left ( \frac{\partial \omega}{\partial z} \right )^{2} = 0,
\end{equation}
where $\omega(t, x, y, z)$ is an electromagnetic wave front, $c$ is the velocity of light in the reference frame $S_{\Omega(1)}$.

An electromagnetic wave front is a characteristics. According to Fock's theorem of characteristics (\cite{FockV1964}, p.\ 432), we obtain the following metric tensor $\eta_{\alpha \beta} = \mathrm{diag} [c^{2},-1,-1,-1]$ of a Minkowski spacetime for vacuum  (\cite{RindlerW1982}, p.\ 57), where $c$ is the speed of light in vacuum.

For convenience, we introduce the following Galilean coordinate system
\begin{equation}\label{notation 400-550}
x^{0} \equiv ct, \quad  \ x^{1} \equiv x, \quad   \  x^{2} \equiv y,  \quad  \  x^{3} \equiv z.
\end{equation}

We will use Greek indices $\alpha, \beta, \mu, \nu$, etc., denote the range $\{0, 1, 2, 3\}$
and use Latin indices $i, j, k$, etc., denote the range $\{1, 2, 3\}$. We will use
Einstein's summation convention, that is, any repeated Greek superscript or subscript appearing in a term of an equation is to be summed from $0$ to $3$.  We introduce the following definition of spacetime interval
\begin{equation}\label{notation 400-100}
ds^{2}=\eta_{\mu \nu}dx^{\mu}dx^{\nu},
\end{equation}
where $\eta_{\mu \nu}$ is the metric tensor of the Minkowski spacetime defined by
$\eta_{\mu \nu} = \mathrm{diag} [1,-1,-1,-1]$.

Suppose that the $\Omega(0)$ substratum is an incompressible viscous fluid. Then, there is no elastic deformations in the fluid and the internal stress states depend on the instantaneous velocity field. Thus, we can choose the reference frame $S_{\Omega(1)}$ as the co-moving coordinate system. The internal energy $U$ is the sum of the internal elastic energy $U_{e}$ and the dissipative energy $U_{d}$, i.e., $U=U_{e}+U_{d}$. Since there is no elastic deformations in the fluid, we have $U_{e}=0$. We introduce the following definition of deviatoric tensor of strain rate $\dot{\gamma}_{j}^{i}$ (\cite{ChenZD2000},p.\ 331)
\begin{equation}\label{strain 400-300}
\dot{\gamma}_{j}^{i}=\dot{S}_{j}^{i}-\dot{S}_{k}^{k}\delta_{j}^{i},
\end{equation}
where $\dot{S}_{j}^{i}$ is the tensor of strain rate, $\dot{S}_{k}^{k}$ is the rate of volume change, $\delta_{j}^{i}$ is the Kronecker delta.

Suppose that the rate of dissipative energy $\dot{U}_{d}$ is the Rayleigh type, then, we have (\cite{ChenZD2000},p.\ 332)
\begin{equation}\label{dissipative 400-400}
\dot{U}_{d}=\mu_{0}\dot{\gamma}_{j}^{i}\dot{\gamma}_{i}^{j},
\end{equation}
where $\mu_{0}$ is the coefficient of viscosity.

Since the $\Omega(0)$ substratum is incompressible, we have $\dot{S}_{k}^{k}=0$. Thus, from Eqs.\ (\ref{dissipative 400-400}) and  Eqs.\ (\ref{strain 400-300}), we have
\begin{equation}\label{dissipative 400-500}
\dot{U}_{d}=\mu_{0}\dot{S}_{j}^{i}\dot{S}_{i}^{j}.
\end{equation}

In the low velocity limit, i.e., $u/c \ll 1$, where $u=|\mbox{\bfseries{u}}|$, the Lagrangian $L_{\Omega(0)}$ for free fields of the $\Omega(0)$ substratum can be written as (\cite{ChenZD2000},p.\ 332)
\begin{equation}\label{Lagrangian 400-600}
L_{\Omega(0)}=\frac{1}{2}\rho_{0}u^{2}+\int_{t_{0}}^{t}\dot{U}_{d}(\dot{S}_{j}^{i})dt,
\end{equation}
where $u=|\mbox{\bfseries{u}}|$, $t_{0}$ is an initial time.

Suppose that the $\Omega(0)$ substratum is a Newtonian fluid and the stress tensor $\sigma_{j}^{i}$ is symmetric, then we have (\cite{LandauLD1987},p.\ 46)
\begin{equation}\label{stress 400-700}
\sigma_{j}^{i}=-p\delta_{j}^{i}+2\mu_{0}\dot{S}_{j}^{i},
\end{equation}
where $p$ is the pressure of the $\Omega(0)$ substratum.

Using Eqs.\ (\ref{stress 400-700}) and Eqs.\ (\ref{dissipative 400-500}), Eqs.\ (\ref{Lagrangian 400-600}) can be written as
\begin{equation}\label{Lagrangian 400-900}
L_{\Omega(0)}=\frac{1}{2}\rho_{0}u^{2}+\int_{t_{0}}^{t}(\sigma_{j}^{i}+p\delta_{j}^{i})\frac{\dot{S}_{i}^{j}}{2}dt,
\end{equation}

For a macroscopic observer, the relaxation time $t_{\varepsilon}$ of the $\Omega(0)$ substratum is so small that the tensor of strain rate $\dot{S}_{j}^{i}$ may be regarded as a slow varying function of time, i.e., $\partial \dot{S}_{j}^{i}/\partial t \ll 1$. Thus, in a small time interval $[t_{0}, t]$, we have $\dot{S}_{j}^{i} \geq 0$, or, $\dot{S}_{j}^{i} \leq 0$. Then, it is possible to choose a value $\bar{\sigma}_{j}^{i}+\bar{p}\delta_{j}^{i}$ of $\sigma_{j}^{i}+p\delta_{j}^{i}$ in the time interval $[t_{0}, t]$ such that Eqs.\ (\ref{Lagrangian 400-900}) can be written as
\begin{equation}\label{Lagrangian 400-1000}
L_{\Omega(0)}=\frac{1}{2}\rho_{0}u^{2}+(\bar{\sigma}_{j}^{i}
+\bar{p}\delta_{j}^{i})\int_{t_{0}}^{t}\frac{\dot{S}_{i}^{j}}{2}dt.
\end{equation}

We introduce the following definition
\begin{equation}\label{potential 400-1100}
\psi_{ij} \stackrel{\bigtriangleup}{=} \int_{t_{0}}^{t}\frac{\dot{S}_{ij}}{2f_{0}}dt,
\end{equation}
where $f_{0}$ is a parameter to be determined.

Using Eqs.\ (\ref{potential 400-1100}), Eqs.\ (\ref{Lagrangian 400-1000}) can be written as
\begin{equation}\label{Lagrangian 400-1200}
L_{\Omega(0)}=\frac{1}{2}\rho_{0}u^{2}+f_{0}\psi_{i}^{j}(\bar{\sigma}_{j}^{i}+\bar{p}\delta_{j}^{i}).
\end{equation}

Since the coefficient of viscosity $\mu_{0}$ of the $\Omega(0)$ substratum may be very small, we introduce the following assumption.
\begin{assumption_my}\label{assumption 400-1300}
In the low velocity limit, i.e., $u/c \ll 1$, where $u=|\mbox{\bfseries{u}}|$, $\mbox{\bfseries{u}}$ is the velocity of the $\Omega(0)$ substratum, we suppose that $\mu_{0} \approx 0$ and we have the following conditions
\begin{equation}\label{potential 400-1310}
\psi_{ij} \approx 0,  \quad  \partial_{\mu} \psi_{ij}\approx 0, \quad  \partial_{\mu}\partial_{\nu} \psi_{ij}\approx 0,
\end{equation}
where
\begin{equation}\label{notation 400-2230}
\partial_{\mu}\equiv\left (\frac{\partial}{\partial x^{0}}, \frac{\partial}{\partial x^{1}}, \frac{\partial}{\partial x^{2}}, \frac{\partial}{\partial x^{3}}\right ).
\end{equation}
\end{assumption_my}

According to the Stokes-Helmholtz resolution theorem, refers to, for
instance, \cite{Eringen1975}, every sufficiently smooth vector field can be decomposed into irrotational
and solenoidal parts. Thus, there exists a scalar function $\varphi$ and a
vector function $\mbox{\upshape\bfseries{R}}$ such that the velocity field
$\mbox{\upshape\bfseries{u}}$ of the $\Omega(0)$ substratum can be represented by \cite{Eringen1975}
\begin{equation}\label{Stokes 400-1400}
\mbox{\upshape\bfseries{u}} = \nabla \varphi + \nabla \times \mbox{\upshape\bfseries{R}},
\end{equation}
where $\nabla \times \varphi =0$, $\nabla \cdot \mbox{\upshape\bfseries{R}} =0$.

We introduce the following definition of a vector function $\vec{\xi}$
\begin{equation}\label{vector 400-1500}
\frac{\partial \vec{\xi}}{\partial (ct)} = \nabla \times \mbox{\upshape\bfseries{R}}.
\end{equation}

Putting Eq.\ (\ref{vector 400-1500}) into Eq.\ (\ref{Stokes 400-1400}), we have
\begin{equation}\label{Stokes 400-1600}
\mbox{\upshape\bfseries{u}} = \nabla \varphi + \frac{\partial \vec{\xi}}{\partial (ct)}.
\end{equation}

Based on Assumption \ref{assumption 400-1300} and using Eq.\ (\ref{definition 200-500}) and Eq.\ (\ref{Stokes 400-1600}), Eq.\ (\ref{Lagrangian 400-1200}) can be written as
\begin{eqnarray}
L_{\Omega(0)}&=&\frac{1}{2}\rho_{0}u^{2}=\frac{1}{2}\rho_{0}\left ( \nabla \varphi + \frac{\partial \vec{\xi}}{\partial (ct)}\right )^{2}\nonumber\\
&=&\frac{1}{2}\rho_{0}\left ( \frac{m_{0}}{\rho_{0}q_{0}}\nabla \Phi + \frac{\partial \vec{\xi}}{\partial (ct)}\right )^{2}. \label{Lagrangian 400-1800}
\end{eqnarray}

We introduce the following definitions
\begin{eqnarray}
&&\psi_{00} = - a_{00}\Phi, \quad \psi_{0i} = \psi_{i0} = a_{0i}\xi_{i}, \label{definition 400-1910}\\
&&\vec{\psi}_{0} = \psi_{01}\mbox{\upshape\bfseries{i}} + \psi_{02}\mbox{\upshape\bfseries{j}} +\psi_{03}\mbox{\upshape\bfseries{k}}.\label{definition 400-1920}
\end{eqnarray}
where $a_{00}>0$ and $a_{0i}>0$ are 4 parameters to be determined.

Eqs.\ (\ref{definition 400-1910}) and Eqs.\ (\ref{potential 400-1100}) have defined a rank 2 symmetric tensor $\psi_{\mu\nu}$. We require that for some special values of $a_{00}$ and $a_{0i}$, Eq.\ (\ref{Lagrangian 400-1800}) can be written as
\begin{eqnarray}
L_{\Omega(0)} &=& \left ( \frac{m_{0}}{q_{0}\sqrt{2\rho_{0}}}\frac{1}{a_{00}}\nabla \psi_{00} - \sqrt{\frac{\rho_{0}}{2}}\frac{\partial (\psi_{0i}/a_{0i})}{\partial (ct)}\mbox{\upshape\bfseries{e}}^{i}\right )^{2} \nonumber\\
& \equiv &\left ( \nabla \psi_{00} - \frac{\partial \vec{\psi}_{0}}{\partial (ct)}\right )^{2}, \label{Lagrangian 400-2000}
\end{eqnarray}
where $\mbox{\upshape\bfseries{e}}^{1} \equiv \mbox{\upshape\bfseries{i}}$, $\mbox{\upshape\bfseries{e}}^{2} \equiv \mbox{\upshape\bfseries{j}}$, $\mbox{\upshape\bfseries{e}}^{3} \equiv \mbox{\upshape\bfseries{k}}$.

Comparing the left- and right-hand parts of Eq.\ (\ref{Lagrangian 400-2000}), we have
\begin{equation}\label{aij 400-2100}
a_{00}=\sqrt{\frac{m_{0}^{2}}{2\rho_{0} q_{0}^{2}}},  \quad  a_{0i} = \sqrt{\frac{\rho_{0}}{2}}.
\end{equation}

In order to construct the Lagrangian $L_{\Omega(0)}$ described in Eq.\ (\ref{Lagrangian 400-2000}) based on the tensorial potential $\psi_{\mu\nu}$, we should consider all the possible products of derivatives of the tensor $\psi_{\mu\nu}$.
If we require that the two tensor indices of $\psi_{\mu\nu}$ are different from each other and the two tensor indices of $\psi_{\mu\nu}$ are different from the derivative index, we have the following two possible products (\cite{FeynmanRP1995}, p.\ 43):
\begin{equation}\label{aij 400-2150}
L_{1} = \partial_{\sigma}\psi_{\mu\nu}\partial^{\sigma}\psi^{\mu\nu}, \quad L_{2} = \partial_{\sigma}\psi_{\mu\nu}\partial^{\mu}\psi^{\sigma\nu},
\end{equation}
where $\psi^{\mu\nu}=\eta^{\mu\lambda}\eta^{\nu\sigma}\psi_{\lambda\sigma}$ is the corresponding contravariant tensor of $\psi_{\mu\nu}$.

If there are two indices of $\psi_{\mu\nu}$ which are equal, or one of the indices of $\psi_{\mu\nu}$ is the same as   the derivative index, we may have the following three possible products (\cite{FeynmanRP1995}, p.\ 43):
\begin{eqnarray}
&&L_{3} = \partial_{\nu}\psi^{\mu\nu}\partial_{\sigma}\psi^{\sigma}_{\ \mu}, \quad L_{4} = \partial^{\mu}\psi_{\mu\nu}\partial^{\nu}\psi, \label{notation 400-2210}\\
&&L_{5} = \partial_{\lambda}\psi\partial^{\lambda}\psi. \label{notation 400-2220}
\end{eqnarray}
where $\psi$ is the trace of $\psi_{\mu\nu}$, i.e., $\psi \equiv \psi_{\lambda}^{\lambda}=\eta_{\alpha\beta}\psi^{\alpha\beta}$,
\begin{equation}\label{notation 400-2240}
\partial^{\mu} \equiv \eta^{\mu\nu}\partial_{\nu}=\left (\frac{\partial}{\partial x^{0}}, -\frac{\partial}{\partial x^{1}}, -\frac{\partial}{\partial x^{2}}, -\frac{\partial}{\partial x^{3}}\right ).
\end{equation}

$L_{3}$ may be omitted because it can be converted to $L_{2}$ by integration by parts (\cite{FeynmanRP1995}, p.\ 43).
\begin{proposition_my}\label{Lagrangian 400-2400}
Suppose that we have the following conditions
\begin{equation}\label{potential 400-2410}
\frac{\partial \psi_{00}}{\partial (ct)}\approx 0, \quad  \frac{\partial \psi_{0i}}{\partial x^{j}}\approx 0.
\end{equation}
If we set
\begin{equation}\label{parameter 400-2415}
c_{1}=\frac{1}{2}, \quad c_{2}=-2, \quad c_{4}=-6, \quad c_{5}=-\frac{3}{2},
\end{equation}
then we have
\begin{equation}\label{potential 400-2420}
c_{1}L_{1}+c_{2}L_{2}+c_{4}L_{4}+c_{5}L_{5} \approx \left ( \nabla \psi_{00} - \frac{\partial \vec{\psi}_{0}}{\partial (ct)}\right )^{2}=\frac{1}{2}\rho_{0}u^{2}.
\end{equation}
\end{proposition_my}
{\bfseries{Proof of Proposition \ref{Lagrangian 400-2400}.}} Based on Eqs.\ (\ref{potential 400-1310}) and Eqs.\ (\ref{potential 400-2410}) and noticing $\psi^{00}=\psi_{00}, \psi^{0i}=-\psi_{0i}$, we have
\begin{equation}\label{Lagrangian 400-2500}
L_{1}\approx -(\nabla \psi_{00})^{2}-2\left ( \frac{\partial \vec{\psi}_{0}}{\partial (ct)}\right )^{2},
\end{equation}
\begin{equation}\label{Lagrangian 400-2600}
L_{2}\approx -2(\nabla \psi_{00})\cdot \frac{\partial \vec{\psi}_{0}}{\partial (ct)}-\left ( \frac{\partial \vec{\psi}_{0}}{\partial (ct)}\right )^{2},
\end{equation}
\begin{eqnarray}
&&L_{3}\approx (\nabla \psi_{00})\cdot \frac{\partial \vec{\psi}_{0}}{\partial (ct)}, \label{Lagrangian 400-2710}\\
&&L_{4}\approx -(\nabla \psi_{00})^{2}.\label{Lagrangian 400-2720}
\end{eqnarray}

Using Eqs.\ (\ref{Lagrangian 400-2500}-\ref{Lagrangian 400-2720}) and Eqs.\ (\ref{parameter 400-2415}), we obtain Eq.\ (\ref{potential 400-2420}). $\Box$

Inspired by W. Thirring \cite{ThirringW1961} and R. P. Feynman (\cite{FeynmanRP1995}, p.\ 43), we introduce the following  assumption.
\begin{assumption_my}\label{assumption 400-2300}
The Lagrangian $L_{\Omega(0)}$ for free fields of the $\Omega(0)$ substratum can be written as
\begin{equation}\label{potential 400-2310}
L_{\Omega(0)}=c_{1}L_{1}+c_{2}L_{2}+c_{4}L_{4}+c_{5}L_{5}+L_{\mathrm{more}},
\end{equation}
where $c_{1}=1/2, \quad c_{2}=-2, \quad c_{4}=-6, \quad c_{5}=-3/2$, $L_{\mathrm{more}}$ denotes those terms involving more than two derivatives of $\psi_{\mu\nu}$.
\end{assumption_my}

\section{Interaction terms of the Lagrangian of a system of the $\Omega(0)$ substratum, the $\Omega(1)$ substratum and matter \label{sec 450}}
In order to derive the field equations, we should explore the possible interaction terms of the Lagrangian of a system of the $\Omega(0)$ substratum, the $\Omega(1)$ substratum and matter. According to Assumption \ref{assumption 400-1300}, the coefficient of viscosity $\mu_{0}$ of the $\Omega(0)$ substratum may be very small. Thus, we may regard the $\Omega(0)$ substratum as an ideal fluid approximately. Then from Eq.\ (\ref{Stokes 400-1400}) we have $\mbox{\bfseries{u}}=\nabla \varphi$. Ignoring the damping force $\rho_{0}Q\mbox{\upshape\bfseries{v}}_s$ in Eq.\ (\ref{force 200-100}) and using $\mbox{\bfseries{u}}=\nabla \varphi$, Eq.\ (\ref{force 200-100}) can be written as
\begin{equation}\label{force 500-2800}
\mbox{\upshape\bfseries{F}}_Q= -\rho_{0}Q\nabla \varphi.
\end{equation}

A particle is modelled as a point sink of the $\Omega(0)$ substratum \cite{WangXS200804,WangXS200810,WangXS2014PhysicsEssays}. Thus, the interaction term of the Lagrangian of a system of the $\Omega(0)$ substratum and a particle can be written as
\begin{equation}\label{interaction 500-2850}
L_{\mathrm{int}}=\rho_{0}Q\varphi.
\end{equation}

Thus, the interaction term of the Lagrangian of a system of the $\Omega(0)$ substratum and continuously distributed  particles can be written as
\begin{equation}\label{interaction 500-2900}
L_{\mathrm{int}}=-\rho_{0}\rho_{s}\varphi.
\end{equation}

Putting Eq.\ (\ref{definition 200-500}) into Eq.\ (\ref{interaction 500-2900}), we have
\begin{equation}\label{interaction 500-3000}
L_{\mathrm{int}}=-\rho_{m}\Phi.
\end{equation}

The $00$ term of the energy-momentum tensor $T^{m}_{\mu\nu}$ of a matter system is $T_{m}^{00}=\rho_{m}c^{2}$. Thus, using Eqs.\ (\ref{definition 400-1910}), Eq.\ (\ref{interaction 500-3000}) can be written as
\begin{equation}\label{interaction 500-3100}
L_{\mathrm{int}}=f_{0}\psi_{00}T_{m}^{00},
\end{equation}
where
\begin{equation}\label{f0 500-3150}
f_{0}=\frac{1}{a_{00}c^{2}}.
\end{equation}

From Eq.\ (\ref{f0 500-3150}), Eq.\ (\ref{aij 400-2100}) and Eq.\ (\ref{constant 200-210}), we have
\begin{equation}\label{f0 500-3160}
f_{0}=\sqrt{\frac{2\rho_{0}q_{0}^{2}}{m_{0}^{2}c^{4}}}=\sqrt{\frac{8\pi \gamma_{N}}{c^{4}}}, \quad \frac{1}{a_{00}^{2}}=8\pi \gamma_{N}.
\end{equation}

Inspired by Eq.\ (\ref{interaction 500-3100}) and Eq.\ (\ref{Lagrangian 400-1200}), we introduce the following assumption.
\begin{assumption_my}\label{assumption 500-3200}
The interaction terms of the Lagrangian of a system of the $\Omega(0)$ substratum, the $\Omega(1)$ substratum and matter can be written in the following form:
\begin{equation}\label{interaction 500-3210}
L_{\mathrm{int}}=f_{0}\psi_{\mu\nu}T_{m}^{\mu\nu}+f_{0}\psi_{\mu\nu}T_{\Omega(1)}^{\mu\nu}+O[(f_{0}\psi_{\mu\nu})^{2}],
\end{equation}
where $T_{m}^{\mu\nu}$ and $T_{\Omega(1)}^{\mu\nu}$ are the contravariant energy-momentum tensors of the system of the matter and the $\Omega(1)$ substratum respectively, $O[(f_{0}\psi_{\mu\nu})^{2}]$ denotes those terms which are small quantities of the order of $(f_{0}\psi_{\mu\nu})^{2}$.
\end{assumption_my}

\section{Derivation of the field equations \label{sec 500}}
Based on Assumptions \ref{assumption 400-2300} and \ref{assumption 500-3200}, the total Lagrangian $L_{\mathrm{tot}}$ of a system of the $\Omega(0)$ substratum, the $\Omega(1)$ substratum and matter can be written as
\begin{eqnarray}
L_{\mathrm{tot}} &= &\frac{1}{2}\partial_{\lambda}\psi_{\mu\nu}\partial^{\lambda}\psi^{\mu\nu} -2\partial_{\lambda}\psi_{\mu\nu}\partial^{\mu}\psi^{\lambda\nu}-6\partial^{\mu}\psi_{\mu\nu}\partial^{\nu}\psi \nonumber\\
&&-\frac{3}{2} \partial_{\lambda}\psi\partial^{\lambda}\psi +L_{\mathrm{more}} +f_{0}\psi_{\mu\nu}(T_{m}^{\mu\nu}+T_{\Omega(1)}^{\mu\nu})\nonumber\\
&&+O[(f_{0}\psi_{\mu\nu})^{2}].\label{Lagrangian 500-3300}
\end{eqnarray}

\begin{theorem_my}\label{field 500-3500}
If we ignore those terms which are small quantities of the order of $(f_{0}\psi_{\mu\nu})^{2}$ and those terms involving more than two derivatives of $\psi_{\mu\nu}$ in Eq.\ (\ref{Lagrangian 500-3300}), i.e., $O[(f_{0}\psi_{\mu\nu})^{2}]$ and $L_{\mathrm{more}}$, then the field equations for the total Lagrangian $L_{\mathrm{tot}}$ in Eq.\ (\ref{Lagrangian 500-3300}) can be written as
\begin{eqnarray}
&&\partial_{\sigma}\partial^{\sigma}\psi_{\alpha\beta}-2(\partial^{\sigma}\partial_{\alpha}\psi_{\beta\sigma}
+\partial^{\sigma}\partial_{\beta}\psi_{\alpha\sigma})
-6(\eta_{\alpha\beta}\partial_{\sigma}\partial_{\lambda}\psi^{\sigma\lambda}\nonumber\\
&&+\partial_{\alpha}\partial_{\beta}\psi)-3\eta_{\alpha\beta}\partial_{\sigma}\partial^{\sigma}\psi
=f_{0}(T^{m}_{\alpha\beta}+T^{\Omega(1)}_{\alpha\beta}).\label{field 500-3510}
\end{eqnarray}
\end{theorem_my}
{\bfseries{Proof of Theorem \ref{field 500-3500}.}}
Starting from the Lagrangian in Eq.\ (\ref{Lagrangian 500-3300}), we have the following Euler-Lagrange equations \cite{BabakSV1999}
\begin{equation}\label{field 500-3400}
\frac{\partial L_{\mathrm{tot}}}{\partial \psi^{\alpha\beta}}-\frac{\partial }{\partial x^{\sigma}}\left ( \frac{\partial L_{\mathrm{tot}}}{\partial (\partial_{\sigma}\psi^{\alpha\beta})}\right )=0.
\end{equation}

We can verify the following results (\cite{FeynmanRP1995}, p.\ 43; \cite{ThirringW1961})
\begin{eqnarray}
\frac{\partial }{\partial x^{\sigma}}\left [ \frac{\partial (\partial_{\lambda}\psi_{\mu\nu}\partial^{\lambda}\psi^{\mu\nu})}{\partial (\partial_{\sigma}\psi^{\alpha\beta})}\right ]&=&2\partial_{\sigma}\partial^{\sigma}\psi_{\alpha\beta}, \label{field 500-3600}\\
\frac{\partial }{\partial x^{\sigma}}\left [ \frac{\partial (\partial_{\lambda}\psi_{\mu\nu}\partial^{\mu}\psi^{\lambda\nu})}{\partial (\partial_{\sigma}\psi^{\alpha\beta})}\right ]&=&\partial^{\sigma}\partial_{\alpha}\psi_{\beta\sigma}+\partial^{\sigma}\partial_{\beta}\psi_{\alpha\sigma}, \label{field 500-3700}\\
\frac{\partial }{\partial x^{\sigma}}\left [ \frac{\partial (\partial^{\mu}\psi_{\mu\nu}\partial^{\nu}\psi)}{\partial (\partial_{\sigma}\psi^{\alpha\beta})}\right ]&=&\partial_{\alpha}\partial_{\beta}\psi+\eta_{\alpha\beta}\partial_{\sigma}\partial_{\lambda}\psi^{\sigma\lambda}, \label{field 500-3800}\\
\frac{\partial }{\partial x^{\sigma}}\left [ \frac{\partial (\partial_{\lambda}\psi\partial^{\lambda}\psi)}{\partial (\partial_{\sigma}\psi^{\alpha\beta})}\right ]&=&2\eta_{\alpha\beta}\partial_{\sigma}\partial^{\sigma}\psi, \label{field 500-3900}\\
\frac{\partial L_{\mathrm{tot}}}{\partial \psi^{\alpha\beta}}&=&f_{0}(T^{m}_{\alpha\beta}+T^{\Omega(1)}_{\alpha\beta}). \label{field 500-3950}
\end{eqnarray}

Putting Eq.\ (\ref{Lagrangian 500-3300}) into Eqs.\ (\ref{field 500-3400}) and using Eqs.\ (\ref{field 500-3600}-\ref{field 500-3950}), we obtain Eqs.\ (\ref{field 500-3510}). $\Box$

For convenience, we introduce the following notation
\begin{eqnarray}
\Psi^{\mu\nu}&=&\partial_{\lambda}\partial^{\lambda}\psi^{\mu\nu}-2\partial_{\lambda}\partial^{\mu}\psi^{\nu\lambda}
-2\partial_{\lambda}\partial^{\nu}\psi^{\mu\lambda}\nonumber\\
&&-6\eta^{\mu\nu}\partial_{\sigma}\partial_{\lambda}\psi^{\sigma\lambda}
-6\partial^{\mu}\partial^{\nu}\psi-3\eta^{\mu\nu}\partial_{\lambda}\partial^{\lambda}\psi.\label{Psi 500-4000}
\end{eqnarray}

Thus, the field equations Eqs.\ (\ref{field 500-3510}) can be written as
\begin{equation}\label{field 500-4100}
\Psi^{\mu\nu}=f_{0}(T_{m}^{\mu\nu}+T_{\Omega(1)}^{\mu\nu}).
\end{equation}

We introduce the following definition of the total energy-momentum tensor $T^{\mu\nu}$ of the system of the matter, the $\Omega(1)$ substratum and the $\Omega(0)$ substratum
\begin{equation}\label{tensor 500-4150}
T^{\mu\nu}=T_{m}^{\mu\nu}+T_{\Omega(1)}^{\mu\nu}+T_{\Omega(0)}^{\mu\nu},
\end{equation}
where $T_{\Omega(0)}^{\mu\nu}$ is the energy-momentum tensor of the $\Omega(0)$ substratum.

Adding the term $f_{0}T_{\Omega(0)}^{\mu\nu}$ on both sides of Eqs.\ (\ref{field 500-4100}) and using Eqs.\ (\ref{tensor 500-4150}), the field equations Eqs.\ (\ref{field 500-4100}) can be written as
\begin{equation}\label{field 500-4200}
\Psi^{\mu\nu}+f_{0}T_{\Omega(0)}^{\mu\nu}=f_{0}T^{\mu\nu}.
\end{equation}

For the total system of matter, the $\Omega(1)$ substratum and the $\Omega(0)$ substratum, the law of conservation of energy and momentum is (\cite{MollerC1955}, p.\ 163; \cite{RindlerW1982}, p.\ 155)
\begin{equation}\label{conservation 500-4300}
\partial_{\mu} T^{\mu\nu}=0.
\end{equation}

Comparing Eqs.\ (\ref{conservation 500-4300}) and Eqs.\ (\ref{field 500-4200}), we have
\begin{equation}\label{conservation 500-4400}
\partial_{\mu} (\Psi^{\mu\nu}+f_{0}T_{\Omega(0)}^{\mu\nu})=0.
\end{equation}

Noticing Eqs.\ (\ref{field 500-3600}-\ref{field 500-3950}), we introduce the following notation (\cite{FeynmanRP1995}, p.\ 43)
\begin{eqnarray}
H^{\mu\nu}&=&f_{1}\partial_{\lambda}\partial^{\lambda}\psi^{\mu\nu}+f_{2}(\partial_{\lambda}\partial^{\mu}\psi^{\nu\lambda}
+\partial_{\lambda}\partial^{\nu}\psi^{\mu\lambda})\nonumber\\
&&+f_{3}(\partial^{\mu}\partial^{\nu}\psi+\eta^{\mu\nu}\partial_{\sigma}\partial_{\lambda}\psi^{\sigma\lambda})\nonumber\\
&&+f_{4}\eta^{\mu\nu}\partial_{\lambda}\partial^{\lambda}\psi,\label{H 500-4410}
\end{eqnarray}
where $f_{i}, i=1,2,3,4$ are 4 arbitrary parameters.

If we require that
\begin{equation}\label{conservation 500-4420}
\partial_{\mu} H^{\mu\nu}=0,
\end{equation}
then, we can verify the following relationships (\cite{FeynmanRP1995}, p.\ 44; \cite{ThirringW1961})
\begin{equation}\label{Feynman 500-4436}
f_{1}+f_{2}=0,  \quad
f_{2}+f_{3}=0,  \quad
f_{3}+f_{4}=0.
\end{equation}

We choose $f_{1}=1$, $f_{2}=-1$, $f_{3}=1$, $f_{4}=-1$ in Eqs.\ (\ref{H 500-4410}) and introduce the following notation
\begin{eqnarray}
\Theta^{\mu\nu}&=&\partial_{\lambda}\partial^{\lambda}\psi^{\mu\nu}-(\partial_{\lambda}\partial^{\mu}\psi^{\nu\lambda}
+\partial_{\lambda}\partial^{\nu}\psi^{\mu\lambda})\nonumber\\
&&+(\partial^{\mu}\partial^{\nu}\psi+\eta^{\mu\nu}\partial_{\sigma}\partial_{\lambda}\psi^{\sigma\lambda})
-\eta^{\mu\nu}\partial_{\lambda}\partial^{\lambda}\psi.\label{Theta 500-4500}
\end{eqnarray}

We can verify the following result (\cite{FeynmanRP1995}, p.\ 44; \cite{ThirringW1961})
\begin{equation}\label{conservation 500-4600}
\partial_{\mu} \Theta^{\mu\nu}=0.
\end{equation}

From Eqs.\ (\ref{conservation 500-4600}) and Eqs.\ (\ref{conservation 500-4400}), we have
\begin{equation}\label{conservation 500-4700}
\partial_{\mu} \left (\frac{1}{f_{0}}\Psi^{\mu\nu}-\frac{b_{0}}{f_{0}}\Theta^{\mu\nu}+T_{\Omega(0)}^{\mu\nu}\right )=0.
\end{equation}
where $b_{0}$ is an arbitrary parameter.

Noticing Eqs.\ (\ref{conservation 500-4700}), it is convenient for us to introduce the following definition of a tensor $T_{\omega}^{\mu\nu}$
\begin{equation}\label{relationship 500-4810}
T_{\omega}^{\mu\nu}=\frac{1}{f_{0}}\Psi^{\mu\nu}-\frac{b_{0}}{f_{0}}\Theta^{\mu\nu}+T_{\Omega(0)}^{\mu\nu},
\end{equation}
where $b_{0}$ is a parameter to be determined.

From Eqs.\ (\ref{conservation 500-4700}), we have $\partial_{\mu} T_{\omega}^{\mu\nu}=0$. In the present stage, we have no idea about the physical meaning of the tensor $T_{\omega}^{\mu\nu}$. Later, once we have determined the value of the parameter $b_{0}$, we may explore the meaning of $T_{\omega}^{\mu\nu}$.
Using Eqs.\ (\ref{relationship 500-4810}), the field equations Eqs.\ (\ref{field 500-4200}) can be written as
\begin{equation}\label{field 500-4970}
\Theta^{\mu\nu}=\frac{f_{0}}{b_{0}}(T^{\mu\nu}-T_{\omega}^{\mu\nu}).
\end{equation}

Now our task is to determine the parameter $b_{0}$ in the field equations (\ref{field 500-4970}).  A natural idea is that the $00$ component of Eqs.\ (\ref{field 500-4970}) reduces to the field equations Eqs.\ (\ref{conservation 200-600}) in the case that the velocity of the $\Omega(0)$ substratum is much smaller than $c$, i.e., in the low velocity limit. Thus, it is necessary for us to introduce an estimation of the value of $T^{\mu\nu}-T_{\omega}^{\mu\nu}$ on the right hand side of Eqs.\ (\ref{field 500-4970}) in the low velocity limit. To this end, we introduce the following speculation about the interaction between the $\Omega(0)$ substratum and the $\Omega(1)$ substratum.

\begin{assumption_my}\label{assumption 500-4850}
In the low velocity limit, i.e., $u/c \ll 1$, where $u=|\mbox{\bfseries{u}}|$, $\mbox{\bfseries{u}}$ is the velocity of the $\Omega(0)$ substratum, the following relationship is valid
\begin{equation}\label{relationship 500-4860}
T_{\Omega(1)}^{\mu\nu} \approx \frac{1}{f_{0}}\Psi^{\mu\nu}-\frac{b_{0}}{f_{0}}\Theta^{\mu\nu},
\end{equation}
where $b_{0}$ is a parameter to be determined.
\end{assumption_my}

Therefore, using Eqs.\ (\ref{field 500-4100}) and Eqs.\ (\ref{relationship 500-4860}), we have the following estimation of $T_{m}^{\mu\nu}$ in the low velocity limit
\begin{equation}\label{relationship 500-4862}
T_{m}^{\mu\nu} \approx \frac{b_{0}}{f_{0}}\Theta^{\mu\nu}.
\end{equation}

Using Eqs.\ (\ref{field 500-4200}), Eqs.\ (\ref{relationship 500-4810}) and Eqs.\ (\ref{relationship 500-4862}), we have the following estimation of $T^{\mu\nu}-T_{\omega}^{\mu\nu}$ in the low velocity limit
\begin{equation}\label{relationship 500-4864}
T^{\mu\nu}-T_{\omega}^{\mu\nu} =  \frac{b_{0}}{f_{0}}\Theta^{\mu\nu} \approx T_{m}^{\mu\nu}.
\end{equation}

\begin{theorem_my}\label{field 500-4950}
Suppose that Assumption \ref{assumption 500-4850} is valid. Then, $b_{0}=-1$ and the field equations (\ref{field 500-3510}) can be written as
\begin{eqnarray}
&&\partial_{\lambda}\partial^{\lambda}\psi^{\mu\nu}-\partial_{\lambda}\partial^{\mu}\psi^{\nu\lambda}
-\partial_{\lambda}\partial^{\nu}\psi^{\mu\lambda}+\partial^{\mu}\partial^{\nu}\psi\nonumber\\
&&+\eta^{\mu\nu}\partial_{\sigma}\partial_{\lambda}\psi^{\sigma\lambda}
-\eta^{\mu\nu}\partial_{\lambda}\partial^{\lambda}\psi=-f_{0}(T^{\mu\nu}-T_{\omega}^{\mu\nu}).\label{field 500-4960}
\end{eqnarray}
\end{theorem_my}
{\bfseries{Proof of Theorem \ref{field 500-4950}.}}  Noticing Eqs.\ (\ref{Theta 500-4500}), the $00$ component of the field equations (\ref{field 500-4970}) is
\begin{eqnarray}
\partial_{\lambda}\partial^{\lambda}\psi^{00}&-&2\partial_{\lambda}\partial^{0}\psi^{0\lambda}
+\partial^{0}\partial^{0}\psi\nonumber\\
&+&\partial_{\sigma}\partial_{\lambda}\psi^{\sigma\lambda}
-\partial_{\lambda}\partial^{\lambda}\psi = \frac{f_{0}}{b_{0}}(T^{00}-T_{\omega}^{00}).\label{field 500-5000}
\end{eqnarray}

Take the trace of the field equations Eqs.\ (\ref{field 500-4970}), we have
\begin{equation}\label{trace 500-5100}
\partial_{\sigma}\partial_{\lambda}\psi^{\sigma\lambda}-\partial_{\lambda}\partial^{\lambda}\psi
=\frac{f_{0}}{2b_{0}}(T-T_{\omega}),
\end{equation}
where $T$ and $T_{\omega}$ are the traces of $T^{\mu\nu}$ and $T_{\omega}^{\mu\nu}$ respectively, i.e., $T \equiv T_{\lambda}^{\lambda}=\eta_{\alpha\beta}T^{\alpha\beta}$, $T_{\omega} \equiv T_{\omega \lambda}^{\ \lambda} =\eta_{\alpha\beta}T_{\omega}^{\alpha\beta}$.

Subtracting Eq.\ (\ref{trace 500-5100}) from Eq.\ (\ref{field 500-5000}), we have
\begin{eqnarray}
\partial_{\lambda}\partial^{\lambda}\psi^{00}&-&2\partial_{\lambda}\partial^{0}\psi^{0\lambda}
+\partial^{0}\partial^{0}\psi\nonumber\\
&=&\frac{f_{0}}{b_{0}}\left (T^{00}-\frac{T}{2}-T_{\omega}^{00}+\frac{T_{\omega}}{2}\right ).\label{field 500-5200}
\end{eqnarray}

If the field is time-independent, then Eq.\ (\ref{field 500-5200}) reduces to
\begin{equation}\label{field 500-5300}
-\nabla^{2}\psi^{00}=\frac{f_{0}}{b_{0}}\left (T^{00}-\frac{T}{2}-T_{\omega}^{00}+\frac{T_{\omega}}{2}\right ).
\end{equation}

According to Eqs.\ (\ref{relationship 500-4864}), we have the following estimations in the low velocity limit
\begin{equation}\label{energy-momentum 500-5400}
T^{00}-T_{\omega}^{00} \approx T_{m}^{00}=\rho_{m}c^{2}, \quad T - T_{\omega} \approx T_{m}\approx \rho_{m}c^{2},
\end{equation}
where $T_{m}$ is the trace of $T_{m}^{\mu\nu}$, i.e., $T_{m} \equiv \eta_{\alpha\beta}T_{m}^{\alpha\beta}$.

Using Eqs.\ (\ref{definition 400-1910}), Eq.\ (\ref{aij 400-2100}), Eq.\ (\ref{f0 500-3160}) and Eqs.\ (\ref{energy-momentum 500-5400}), Eq.\ (\ref{field 500-5300}) can be written as
\begin{equation}\label{field 500-5500}
\nabla^{2}\Phi=\frac{1}{b_{0}}4\pi\gamma_{N}\rho_{m}.
\end{equation}

Comparing Eq.\ (\ref{field 500-5500}) and Eq.\ (\ref{conservation 200-600}), we obtain $b_{0}=-1$.
Therefore, using Eqs.\ (\ref{Theta 500-4500}) and $b_{0}=-1$, the field equations Eqs.\ (\ref{field 500-4970}) can be written as Eqs.\ (\ref{field 500-4960}).  $\Box$

Now we discuss the physical meaning of $T_{\omega}^{\mu\nu}$. Noticing Eqs.\ (\ref{relationship 500-4860}) and Eqs.\ (\ref{relationship 500-4810}), we have the following estimation in the low velocity limit
\begin{equation}\label{estimation 500-5550}
T_{\omega}^{\mu\nu} \approx T_{\Omega(1)}^{\mu\nu}+T_{\Omega(0)}^{\mu\nu}.
\end{equation}

For convenience, we may call $T_{\Omega}^{\mu\nu}\stackrel{\bigtriangleup}{=}T_{\Omega(1)}^{\mu\nu}+T_{\Omega(0)}^{\mu\nu}$ the contravariant energy-momentum tensor of vacuum. From Eqs.\ (\ref{estimation 500-5550}), we see that the tensor $T_{\omega}^{\mu\nu}$ is an estimation of $T_{\Omega}^{\mu\nu}$ when the velocity $u$ of the $\Omega(0)$ substratum is small comparing to $c$. Thus, we may call $T_{\omega}^{\mu\nu}$ the contravariant energy-momentum tensor of vacuum in the low velocity limit. We can verify that the field equations Eqs.\ (\ref{field 500-4960}) is invariant under the following gauge transformation (\cite{FeynmanRP1995}, p.\ 45; \cite{ThirringW1961})
\begin{equation}\label{definition 500-5750}
\psi^{\mu\nu} \rightarrow \psi^{\mu\nu} + \partial^{\mu}\Lambda^{\nu} +\partial^{\nu}\Lambda^{\mu},
\end{equation}
where $\Lambda^{\mu}$ is an arbitrary vector field.

We introduce the following definition
\begin{equation}\label{definition 500-6100}
\phi^{\mu\nu}=\psi^{\mu\nu}-\frac{1}{2}\eta^{\mu\nu}\psi.
\end{equation}

Using Eqs.\ (\ref{definition 500-6100}), the field equations (\ref{field 500-4960}) can be written as
\begin{eqnarray}
\partial_{\lambda}\partial^{\lambda}\phi^{\mu\nu}&-&\partial_{\lambda}\partial^{\mu}\phi^{\nu\lambda}
-\partial_{\lambda}\partial^{\nu}\phi^{\mu\lambda}\nonumber\\
&+&\eta^{\mu\nu}\partial_{\sigma}\partial_{\lambda}\phi^{\sigma\lambda}= -f_{0}(T^{\mu\nu}-T_{\omega}^{\mu\nu}).\label{field 500-5755}
\end{eqnarray}

We introduce the following Hilbert gauge condition \cite{ThirringW1961}
\begin{equation}\label{Hilbert 500-5800}
\partial_{\mu}\left ( \psi^{\mu\nu}-\frac{1}{2}\eta^{\mu\nu}\psi \right )=0.
\end{equation}

Using Eqs.\ (\ref{definition 500-6100}), the Hilbert gauge condition Eqs.\ (\ref{Hilbert 500-5800}) simplifies to
\begin{equation}\label{Hilbert 500-6200}
\partial_{\mu}\phi^{\mu\nu}=0.
\end{equation}

Applying Eqs.\ (\ref{Hilbert 500-6200}) in Eqs.\ (\ref{field 500-5755}), we obtain the following proposition \cite{ThirringW1961}.
\begin{proposition_my}\label{field 500-5760}
If we impose the Hilbert gauge condition Eqs.\ (\ref{Hilbert 500-5800}) on the fields, then, the field equations Eqs.\ (\ref{field 500-4960}) simplifies to
\begin{equation}\label{field 500-5900}
\partial_{\lambda}\partial^{\lambda}\left ( \psi^{\mu\nu}-\frac{1}{2}\eta^{\mu\nu}\psi \right )=-f_{0}(T^{\mu\nu}-T_{\omega}^{\mu\nu}).
\end{equation}
\end{proposition_my}

If the tensorial potential $\psi^{\mu\nu}$ does not satisfy the Hilbert gauge condition Eqs.\ (\ref{Hilbert 500-5800}), then we can always construct a new tensorial potential $\bar{\psi}^{\mu\nu}$ by the following gauge transformation \cite{ThirringW1961}
\begin{equation}\label{definition 500-6000}
\bar{\psi}^{\mu\nu} = \psi^{\mu\nu} + \partial^{\mu}\Lambda^{\nu} +\partial^{\nu}\Lambda^{\mu},
\end{equation}
such that the new tensorial potential $\bar{\psi}^{\mu\nu}$ does satisfy the Hilbert gauge condition Eqs.\ (\ref{Hilbert 500-5800}).

Using Eqs.\ (\ref{definition 500-6100}), the field equations Eqs.\ (\ref{field 500-5900}) can be written as
\begin{equation}\label{field 500-6300}
\partial_{\lambda}\partial^{\lambda}\phi^{\mu\nu}=-f_{0}(T^{\mu\nu}-T_{\omega}^{\mu\nu}).
\end{equation}

The field equation Eqs.\ (\ref{field 500-6300}) can also be written as
\begin{equation}\label{field 500-6400}
\eta^{\alpha\beta}\frac{\partial^{2}\phi^{\mu\nu}}{\partial x^{\alpha} \partial x^{\beta}}=-f_{0}(T^{\mu\nu}-T_{\omega}^{\mu\nu}).
\end{equation}

We noticed that the tensorial field equations Eqs.\ (\ref{field 500-6400}) is similar to the wave equations of electromagnetic fields.

\section{Construction of a tensorial potential in inertial coordinate systems \label{sec 700}}
The existence of the $\Omega(1)$ substratum allows us to introduce the following definition of inertial coordinate systems.

\begin{definition_my}\label{inertia 700-60}
If a coordinates system $S$ is static or moving with a constant velocity relative to the reference frame $S_{\Omega(1)}$, then, we call such a coordinates system as an inertial coordinate system.
\end{definition_my}

The field equations Eqs.\ (\ref{field 500-5755}) and Eqs.\ (\ref{field 500-5900}) are valid in the reference frame $S_{\Omega(1)}$. We will explore the possibility of constructing a tensorial potential in an arbitrary inertial system $S'$. In an inertial coordinate system $S$, an arbitrary event is characterized by the four space-time coordinates $(t, x, y, z)$. In an inertial system $S'$, this event is characterized by four other coordinates $(t', x', y', z')$. We assume that the origins of the Cartesian coordinates in the two inertial systems $S$ and $S'$ coincide at the time $t=t'=0$. Then, the connections between these space-time coordinates are given by a homogeneous linear transformation keeping the quantity $s^{2}=c^{2}t^{2}-x^{2}-y^{2}-z^{2}$ invariant, i.e., (\cite{MollerC1955}, p.\ 92)
\begin{equation}\label{invariant 700-100}
s^{2}=c^{2}t^{2}-x^{2}-y^{2}-z^{2}=c^{2}t'^{2}-x'^{2}-y'^{2}-z'^{2}=s'^{2}.
\end{equation}

We introduce the following two coordinate systems
\begin{eqnarray}
&&x^{0} = ct,  \quad  x^{1} = x,   \quad   x^{2} = y,   \quad   x^{3} = z, \nonumber\\
&&x'^{0} = ct',  \quad  x'^{1} = x',   \quad   x'^{2} = y',   \quad   x'^{3} = z'. \label{coordinate 700-200}
\end{eqnarray}

The homogeneous linear transformation keeping the quantity $s^{2}$ invariant, which is usually called the Lorentz transformation, can be written as (\cite{OhanianHC2013}, p.\ 57; \cite{MollerC1955}, p.\ 92)
\begin{equation}\label{Lorentz 700-300}
x'^{\mu}=\alpha^{\mu}_{\  \nu}x^{\nu},
\end{equation}
where $\alpha^{\mu}_{\  \nu}$ are coefficients depend only on the angles between the spatial axes in the two inertial systems $S$ and $S'$ and on the relative velocity of $S$ and $S'$.

Applying the standard methods in theory of special relativity \cite{MollerC1955}, we have the following results.
\begin{proposition_my}\label{potential 700-400}
Suppose that the field equations Eqs.\ (\ref{field 500-6300}) is valid in the the reference frame $S_{\Omega(1)}$. Then, in an arbitrary inertial system $S'$, there exists a symmetric tensor $\phi'_{\mu\nu}$ satisfies the following wave equation
\begin{equation}\label{field 700-410}
\partial'_{\lambda}\partial'^{\lambda}\phi'^{\mu\nu}=-f_{0}(T'^{\mu\nu}-T_{\omega}'^{\mu\nu}),
\end{equation}
where $T'^{\mu\nu}$ and $T_{\omega}'^{\mu\nu}$ are corresponding tensors of $T^{\mu\nu}$ and $T_{\omega}^{\mu\nu}$ in the arbitrary inertial coordinate system $S'$ respectively.
\end{proposition_my}

\begin{proposition_my}\label{potential 700-1100}
Suppose that the field equations Eqs.\ (\ref{field 500-5755}) is valid in the reference frame $S_{\Omega(1)}$. Then, in an arbitrary inertial system $S'$, there exists a symmetric tensor $\phi'_{\mu\nu}$ satisfies the following field equation
\begin{eqnarray}
\partial'_{\lambda}\partial'^{\lambda}\phi'^{\mu\nu}&-&\partial'_{\lambda}\partial'^{\mu}\phi'^{\nu\lambda}
-\partial'_{\lambda}\partial'^{\nu}\phi'^{\mu\lambda}\nonumber\\
&+&\eta^{\mu\nu}\partial'_{\sigma}\partial'_{\lambda}\phi'^{\sigma\lambda}= -f_{0}(T'^{\mu\nu}-T_{\omega}'^{\mu\nu}).\label{field 700-1110}
\end{eqnarray}
\end{proposition_my}

\section{The equations of motion of a point particle in a gravitational field and introduction of an effective  Riemannian spacetime \label{sec 750}}
In this section, we study the equations of motion of a free point particle in a gravitational field. The Lagrangian of a free point particle can be written as (\cite{FeynmanRP1995}, p.\ 57;\cite{ThirringW1961})
\begin{equation}\label{Lagrangian 750-100}
L_{0} = \frac{1}{2}m\frac{dx^{\mu}}{d\tau}\frac{dx_{\mu}}{d\tau}=\frac{1}{2}mu^{\mu}u_{\mu}=\frac{1}{2}m\eta_{\mu\nu}u^{\mu}u^{\nu},
\end{equation}
where $m$ is the rest mass of the point particle, $d\tau \equiv \frac{1}{c}\sqrt{dx^{\mu}dx_{\mu}}$ is the infinitesimal proper time interval, $u^{\mu}\equiv dx^{\mu}/d\tau$.

Suppose that $T_{\Omega(1)}^{\mu\nu}\approx 0$. Ignoring those higher terms $O[(f_{0}\psi_{\mu\nu})^{2}]$ in Eq.\ (\ref{interaction 500-3210}), the interaction term of the Lagrangian of a system of the $\Omega(0)$ substratum, the $\Omega(1)$ substratum and the point particle can be written in the following form (\cite{FeynmanRP1995}, p.\ 57;\cite{ThirringW1961})
\begin{equation}\label{interaction 750-200}
L_{\mathrm{int}}=f_{0}\psi_{\mu\nu}mu^{\mu}u^{\nu}.
\end{equation}

Using Eq.\ (\ref{interaction 750-200}) and Eq.\ (\ref{Lagrangian 750-100}), the total Lagrangian $L_{\mathrm{tot}}$ of a system of the $\Omega(0)$ substratum, the $\Omega(1)$ substratum and the point particle can be written as (\cite{FeynmanRP1995}, p.\ 57)
\begin{equation}\label{Lagrangian 750-300}
L_{\mathrm{tot}}=L_{0}+L_{\mathrm{int}}=\frac{1}{2}mu^{\mu}u_{\mu}+f_{0}\psi_{\mu\nu}mu^{\mu}u^{\nu}.
\end{equation}

The Euler-Lagrange equations for the total Lagrangian $L_{\mathrm{tot}}$ can be written as (\cite{OhanianHC2013},p.\ 111)
\begin{equation}\label{motion 750-1000}
\frac{d}{d\tau}\left [ (\eta_{\mu\nu}+2f_{0}\psi_{\mu\nu})\frac{dx^{\nu}}{d\tau}\right ] - f_{0}\frac{\partial \psi_{\alpha\beta}}{\partial x^{\mu}}\frac{dx^{\alpha}}{d \tau}\frac{dx^{\beta}}{d\tau} = 0.
\end{equation}

We notice that the equations of motion (\ref{motion 750-1000}) of a point particle in gravitational field are similar to the equations of a geodesic line (\ref{motion 750-1200}) in a Riemannian spacetime. Thus, it is natural for us to introduce the following definition of a metric tensor $g_{\mu\nu}$ of a Riemannian spacetime (\cite{FeynmanRP1995}, p.\ 57)
\begin{equation}\label{definition 750-1100}
g_{\mu\nu}=\eta_{\mu\nu}+2f_{0}\psi_{\mu\nu}.
\end{equation}

Then, the equations of motion Eqs.\ (\ref{motion 750-1000}) can be written as (\cite{FeynmanRP1995}, p.\ 58)
\begin{equation}\label{motion 750-1200}
\frac{d}{d\tau_{g}}\left (g_{\mu\nu}\frac{dx^{\nu}}{d\tau_{g}}\right )= \frac{1}{2}\frac{\partial g_{\alpha\beta}}{\partial x^{\mu}}\frac{dx^{\alpha}}{d \tau_{g}}\frac{dx^{\beta}}{d\tau_{g}},
\end{equation}
where $\tau_{g}$ is the proper time interval in the Riemannian spacetime with a metric tensor $g_{\mu\nu}$.

Eqs.\ (\ref{motion 750-1200}) represent a geodesic line in a Riemannian spacetime with a metric tensor $g_{\mu\nu}$, which can also be written as (\cite{LiuL2004}, p.\ 51)
\begin{equation}\label{motion 750-1250}
\frac{d^{2}x^{\mu}}{d\tau_{g}^{2}}+\Gamma_{\nu\sigma}^{\mu}\frac{dx^{\nu}}{d \tau_{g}}\frac{dx^{\sigma}}{d\tau_{g}}= 0,
\end{equation}
where
\begin{equation}\label{Christoffel 750-1260}
\Gamma_{\alpha\beta}^{\nu}\stackrel{\bigtriangleup}{=}\frac{1}{2}g^{\mu\nu}
\left ( \frac{\partial g_{\mu\alpha}}{\partial x^{\beta}}
+\frac{\partial g_{\mu\beta}}{\partial x^{\alpha}}-\frac{\partial g_{\alpha\beta}}{\partial x^{\mu}}\right )
\end{equation}
are the Christoffel symbols.

Thus, we find that the equations of motion (\ref{motion 750-1000}) of a point particle in gravitational field are approximately a geodesic line described in Eqs.\ (\ref{motion 750-1250}) in a Riemannian spacetime with a metric tensor $g_{\mu\nu}$.

According to Assumption \ref{assumption 300-200}, the particles that constitute the $\Omega(1)$ substratum are sinks in the $\Omega(0)$ substratum. Thus, the movements of the $\Omega(1)$ substratum in gravitational field will be different from the Maxwell's equations. We notice that the equations of motion of a point particle in gravitational field (\ref{motion 750-1200}) are generalizations of the equations of motion of a point particle in vacuum free of gravitational field. The law of propagation of an electromagnetic wave front in vacuum free of gravitational field is Eqs.\ (\ref{front 400-400}). Thus, the law of propagation of an electromagnetic wave front in gravitational field may be a kind of generalization of Eq.\ (\ref{front 400-400}). Therefore, we introduce the following assumption.

\begin{assumption_my}\label{assumption 750-1300}
To first order of $f_{0}\psi_{\mu\nu}$, the law of propagation of an electromagnetic wave front $\omega(x^{0}, x^{1}, x^{2}, x^{3})$ in gravitational field is
\begin{equation}\label{front 750-1310}
g_{\mu \nu}\frac{\partial \omega}{\partial x^{\mu}}\frac{\partial \omega}{\partial x^{\nu}}= 0,
\end{equation}
where $\omega(x^{0}, x^{1}, x^{2}, x^{3})$ is the electromagnetic wave front, $g_{\alpha \beta}$ is the metric tensor defined in Eqs.\ (\ref{definition 750-1100}).
\end{assumption_my}

The measurements of spacetime intervals are carried out using light rays and point particles, which are only subject to inertial force and gravitation. Thus, according to Eqs.\ (\ref{motion 750-1200}) and Eq.\ (\ref{front 750-1310}), the physically observable metric of spacetime, to first order of $f_{0}\psi_{\mu\nu}$, is $g_{\mu\nu}$. Thus, the initial flat background spacetime with metric $\eta_{\mu\nu}$ is no longer physically observable \cite{ThirringW1961}.

 If we can further derive the Einstein's equations (\ref{Einstein 100-100}) using the definition (\ref{definition 750-1100}) of a metric tensor $g_{\mu\nu}$ of a Riemannian spacetime, then, we may provide a geometrical interpretation of Einstein's theory of gravitation based on the theory of vacuum mechanics \cite{WangXS200804,WangXS200810,WangXS2014PhysicsEssays}. This is the task of the next section.

\section{Derivation of generalized Einstein equations in inertial coordinate systems \label{sec 800}}
\begin{definition_my}\label{Einstein 800-2800}
The Einstein tensor $G_{\mu\nu}$ is defined by
\begin{equation}\label{Einstein 800-2810}
G_{\mu\nu} \stackrel{\bigtriangleup}{=} R_{\mu\nu}-\frac{1}{2}g_{\mu\nu}R,
\end{equation}
where $g_{\mu\nu}$ is a metric tensor of a Riemannian spacetime,
$R_{\mu\nu}$ is the Ricci tensor, $R\stackrel{\bigtriangleup}{=}g^{\mu\nu}R_{\mu\nu}$, $g^{\mu\nu}$ is the corresponding contravariant tensor of $g_{\mu\nu}$ such that $g_{\mu\lambda}g^{\lambda\nu}=\delta_{\mu}^{\nu}=g_{\mu}^{\nu}$ (\cite{LiuL2004}, p.\ 40).
\end{definition_my}

According to the geometrical interpretation of some theories of gravitation in flat spacetime \cite{ThirringW1961}, the physically observable metric $g_{\mu\nu}$ of spacetime in Eqs.\ (\ref{definition 750-1100}) can be written as
\begin{equation}\label{metric 800-2990}
g^{\mu\nu}=\eta^{\mu\nu}-2f_{0}\psi^{\mu\nu} + O[(f_{0}\psi^{\mu\nu})^{2}].
\end{equation}

Following the clue showed in Eqs.\ (\ref{metric 800-2990}) and noticing the methods of S. N. Gupta \cite{GuptaSN1952} and W. Thirring \cite{ThirringW1961}, we introduce the following definition of a metric tensor of a Riemannian spacetime.
\begin{definition_my}\label{metric 800-3000}
\begin{equation}\label{metric 800-3010}
\tilde{g}^{\mu\nu}\stackrel{\bigtriangleup}{=}\sqrt{-g_{0}}g^{\mu\nu} \stackrel{\bigtriangleup}{=}\eta^{\mu\nu}-2f_{0}\phi^{\mu\nu},
\end{equation}
where $g_{0} = \mathrm{Det} \ g_{\mu\nu}$.
\end{definition_my}

We have the following expansion of the contravariant metric tensor $g^{\mu\nu}$ \cite{GuptaSN1952}
\begin{eqnarray}
g^{\mu\nu}&=&\eta^{\mu\nu}-2f_{0}\phi^{\mu\nu}
+f_{0}\eta^{\mu\nu}\eta_{\alpha\beta}\phi^{\alpha\beta}\nonumber\\
&&-2f_{0}^{2}\eta_{\alpha\beta}\phi^{\alpha\beta}\phi^{\mu\nu}
+f_{0}^{2}\eta^{\mu\nu}\eta_{\alpha\sigma}\eta_{\beta\lambda}\phi^{\alpha\beta}\phi^{\lambda\sigma}\nonumber\\
&&+\frac{1}{2}f_{0}^{2}\eta^{\mu\nu}\eta_{\alpha\beta}\eta_{\lambda\sigma}\phi^{\alpha\beta}\phi^{\lambda\sigma}
+O[(f_{0}\phi^{\mu\nu})^{3}].
\label{expansion 800-3900}
\end{eqnarray}

\begin{definition_my}\label{weak 800-3700}
If $\phi^{\mu\nu}$ and their first and higher derivatives satisfy the following conditions
\begin{eqnarray}
&& \left | 2f_{0}\phi^{\mu\nu} \right |\ll 1, \label{weak 800-3710} \\
&&\left | \frac{\partial^{n} (2f_{0}\phi^{\mu\nu})}{\partial (x^{\alpha})^{n}} \right | \ll 1, n=1, 2, 3, \cdots\label{weak 800-3720}
\end{eqnarray}
then we call this filed $\phi^{\mu\nu}$ weak.
\end{definition_my}

For weak fields, $\psi \approx \phi \approx 0$. Thus, $\phi^{\mu\nu}=\psi^{\mu\nu}-1/2 \cdot \eta^{\mu\nu}\psi \approx \psi^{\mu\nu}$. From Eqs.\ (\ref{expansion 800-3900}), we see that the definition (\ref{metric 800-3010}) is compatible with Eqs.\ (\ref{metric 800-2990}).

\begin{theorem_my}\label{field 800-3050}
Suppose that Assumption \ref{assumption 500-4850} is valid. Then, in an arbitrary inertial coordinate system $S_{i}$,  we have the following field equations
\begin{eqnarray}
G^{\mu\nu}&-&\frac{1}{2g_{0}}\left (\sqrt{-g_{0}}g^{\alpha\beta}-\eta^{\alpha\beta}\right )\frac{\partial^{2}(\sqrt{-g_{0}}g^{\mu\nu})}{\partial x^{\alpha} \partial x^{\beta}}\nonumber\\
&-&\frac{\sqrt{-g_{0}}}{2g_{0}}(\partial_{\lambda}\partial^{\mu}g^{\nu\lambda}
+\partial_{\lambda}\partial^{\nu}g^{\mu\lambda}
-\eta^{\mu\nu}\partial_{\sigma}\partial_{\lambda}g^{\sigma\lambda})\nonumber\\
&-&\Pi^{\mu,\alpha\beta}\Pi_{\alpha\beta}^{\nu}+\frac{1}{2}y^{\mu}y^{\nu}
-\frac{1}{2}g^{\mu\nu}(L+B)\nonumber\\
&+&B^{\mu\nu}=\frac{f_{0}^{2}}{g_{0}}(T^{\mu\nu}-T_{\omega}^{\mu\nu}),\label{Einstein 800-3055}
\end{eqnarray}
where $T^{\mu\nu}$ is the contravariant total energy-momentum tensor of the system of the matter, the $\Omega(1)$ substratum and the $\Omega(0)$ substratum in the inertial coordinate system $S_{i}$, $T_{\omega}^{\mu\nu}$ is the contravariant energy-momentum tensor of vacuum in the low velocity limit in $S_{i}$,
\begin{equation}\label{Fock 800-2920}
\Pi^{\mu,\alpha\beta}\stackrel{\bigtriangleup}{=}\frac{1}{2g_{0}}
\left ( \tilde{g}^{\alpha\lambda}\frac{\partial \tilde{g}^{\mu\beta}}{\partial x^{\lambda}}
+\tilde{g}^{\beta\lambda}\frac{\partial \tilde{g}^{\mu\alpha}}{\partial x^{\lambda}}-\tilde{g}^{\mu\lambda}\frac{\partial \tilde{g}^{\alpha\beta}}{\partial x^{\lambda}}\right ),
\end{equation}
\begin{equation}\label{Fock 800-2930}
\Pi_{\alpha\beta}^{\nu}\stackrel{\bigtriangleup}{=}g_{\alpha\lambda}g_{\beta\sigma}\Pi^{\nu,\lambda\sigma},
\end{equation}
\begin{equation}\label{Fock 800-2940}
\Gamma^{\alpha}\stackrel{\bigtriangleup}{=}g^{\sigma\lambda}\Gamma_{\sigma\lambda}^{\alpha},
\end{equation}
\begin{equation}\label{Fock 800-2950}
\Gamma^{\mu\nu} \stackrel{\bigtriangleup}{=} \frac{1}{2}
\left ( g^{\mu\alpha}\frac{\partial \Gamma^{\nu}}{\partial x^{\alpha}}
 + g^{\nu\alpha}\frac{\partial \Gamma^{\mu}}{\partial x^{\alpha}}
- \frac{\partial g^{\mu\nu}}{\partial x^{\alpha}}\Gamma^{\alpha}\right ),
\end{equation}
\begin{equation}\label{Fock 800-2960}
y_{\beta}\stackrel{\bigtriangleup}{=}\frac{\partial(\lg\sqrt{-g_{0}})}{\partial x^{\beta}}, \ y^{\alpha}\stackrel{\bigtriangleup}{=}g^{\alpha\beta}y_{\beta},
\end{equation}
\begin{equation}\label{Fock 800-2970}
L \stackrel{\bigtriangleup}{=} -\frac{1}{2}\Gamma_{\alpha\beta}^{\nu}\frac{\partial  g^{\alpha\beta}}{\partial x^{\nu}}
-\Gamma^{\alpha}\frac{\partial(\lg\sqrt{-g_{0}})}{\partial x^{\alpha}},
\end{equation}
\begin{equation}\label{Fock 800-2980}
B^{\mu\nu}\stackrel{\bigtriangleup}{=}\Gamma^{\mu\nu}+\frac{1}{2}(y^{\mu}\Gamma^{\nu}+y^{\nu}\Gamma^{\mu}),\ B\stackrel{\bigtriangleup}{=}g_{\mu\nu}B^{\mu\nu}.
\end{equation}
\end{theorem_my}
{\bfseries{Proof of Theorem \ref{field 800-3050}.}}
According to a theorem of V. Fock (\cite{FockV1964}, p.\ 429), the contravariant Einstein tensor $G^{\mu\nu}$ can be written as
\begin{eqnarray}
G^{\mu\nu}&=&\frac{1}{2g_{0}}\tilde{g}^{\alpha\beta}\frac{\partial^{2} \tilde{g}^{\mu\nu}}{\partial x_{\alpha} \partial x_{\beta}}
+\Pi^{\mu,\alpha\beta}\Pi_{\alpha\beta}^{\nu}-\frac{1}{2}y^{\mu}y^{\nu} \nonumber\\
&&+\frac{1}{2}g^{\mu\nu}(L+B)-B^{\mu\nu}.\label{Fock 800-2910}
\end{eqnarray}

Applying Eqs.\ (\ref{metric 800-3010}), Eqs.\ (\ref{Fock 800-2910}) can be written as
\begin{eqnarray}
G^{\mu\nu}&=&\frac{1}{2g_{0}}\left (\sqrt{-g_{0}}g^{\alpha\beta}-\eta^{\alpha\beta}\right )\frac{\partial^{2}(-2f_{0}\phi^{\mu\nu})}{\partial x^{\alpha} \partial x^{\beta}} \nonumber\\
&&-\frac{f_{0}}{g_{0}}\eta^{\alpha\beta}\frac{\partial^{2}\phi^{\mu\nu}}{\partial x^{\alpha} \partial x^{\beta}} +\Pi^{\mu,\alpha\beta}\Pi_{\alpha\beta}^{\nu}\nonumber\\
&&-\frac{1}{2}y^{\mu}y^{\nu}+\frac{1}{2}g^{\mu\nu}(L+B)-B^{\mu\nu}.\label{Einstein 800-3060}
\end{eqnarray}

Noticing Eqs.\ (\ref{metric 800-3010}), the field equations Eqs.\ (\ref{field 700-1110}) can be written as
\begin{eqnarray}
\eta^{\alpha\beta}\frac{\partial^{2}\phi^{\mu\nu}}{\partial x^{\alpha} \partial x^{\beta}}=&-& \frac{\sqrt{-g_{0}}}{2f_{0}}\left ( \partial_{\lambda}\partial^{\mu}g^{\nu\lambda}
+\partial_{\lambda}\partial^{\nu}g^{\mu\lambda} \right .\nonumber\\
&-&\left . \eta^{\mu\nu}\partial_{\sigma}\partial_{\lambda}g^{\sigma\lambda} \right ) -f_{0}(T^{\mu\nu}-T_{\omega}^{\mu\nu}).\label{field 800-3070}
\end{eqnarray}

Using Eqs.\ (\ref{metric 800-3010}) and Eqs.\ (\ref{field 800-3070}), Eqs.\ (\ref{Einstein 800-3060}) can be written as
\begin{eqnarray}
G^{\mu\nu}&=&\frac{1}{2g_{0}}\left (\sqrt{-g_{0}}g^{\alpha\beta}-\eta^{\alpha\beta}\right )\frac{\partial^{2}(\sqrt{-g_{0}}g^{\mu\nu})}{\partial x^{\alpha} \partial x^{\beta}} \nonumber\\
&&+\frac{\sqrt{-g_{0}}}{2g_{0}}(\partial_{\lambda}\partial^{\mu}g^{\nu\lambda}
+\partial_{\lambda}\partial^{\nu}g^{\mu\lambda}
-\eta^{\mu\nu}\partial_{\sigma}\partial_{\lambda}g^{\sigma\lambda})\nonumber\\
&&+\frac{f_{0}^{2}}{g_{0}}(T^{\mu\nu}-T_{\omega}^{\mu\nu})+\Pi^{\mu,\alpha\beta}\Pi_{\alpha\beta}^{\nu}
-\frac{1}{2}y^{\mu}y^{\nu}\nonumber\\
&&+\frac{1}{2}g^{\mu\nu}(L+B)-B^{\mu\nu}.\label{Einstein 800-3080}
\end{eqnarray}

Eqs.\ (\ref{Einstein 800-3080}) can be written as Eqs.\ (\ref{Einstein 800-3055}). $\Box$

Eqs.\ (\ref{Einstein 800-3055}) have the same form in all inertial coordinate systems. Eqs.\ (\ref{Einstein 800-3055}) is one of the main results in this manuscript. We need to further study the relationship between Eqs.\ (\ref{Einstein 800-3055}) and the Einstein field equations Eqs.\ (\ref{Einstein 100-100}).

\begin{theorem_my}\label{field 800-3100}
If we impose the Hilbert gauge Eqs.\ (\ref{Hilbert 500-5800}) on the fields, then in an arbitrary inertial coordinate system $S_{i}$ we have the following field equations
\begin{eqnarray}
G^{\mu\nu}&-&\frac{1}{2g_{0}}\left (\sqrt{-g_{0}}g^{\alpha\beta}-\eta^{\alpha\beta}\right )\frac{\partial^{2}(\sqrt{-g_{0}}g^{\mu\nu})}{\partial x^{\alpha} \partial x^{\beta}} \nonumber\\
&-&\Pi^{\mu,\alpha\beta}\Pi_{\alpha\beta}^{\nu}+\frac{1}{2}y^{\mu}y^{\nu}-\frac{1}{2}g^{\mu\nu}(L+B)\nonumber\\
&+&B^{\mu\nu}=\frac{f_{0}^{2}}{g_{0}}(T^{\mu\nu}-T_{\omega}^{\mu\nu}).\label{Einstein 800-3110}
\end{eqnarray}
\end{theorem_my}
{\bfseries{Proof of Theorem \ref{field 800-3100}.}} Using Eqs.\ (\ref{metric 800-3010}) and Eqs.\ (\ref{field 700-410}), Eqs.\ (\ref{Einstein 800-3060}) can be written as
\begin{eqnarray}
G^{\mu\nu}&=&\frac{1}{2g_{0}}\left (\sqrt{-g_{0}}g^{\alpha\beta}-\eta^{\alpha\beta}\right )\frac{\partial^{2}(\sqrt{-g_{0}}g^{\mu\nu})}{\partial x^{\alpha} \partial x^{\beta}} \nonumber\\
&&+\frac{f_{0}^{2}}{g_{0}}(T^{\mu\nu}-T_{\omega}^{\mu\nu}) +\Pi^{\mu,\alpha\beta}\Pi_{\alpha\beta}^{\nu}-\frac{1}{2}y^{\mu}y^{\nu}\nonumber\\
&&+\frac{1}{2}g^{\mu\nu}(L+B)-B^{\mu\nu}.\label{Einstein 800-3300}
\end{eqnarray}

Eqs.\ (\ref{Einstein 800-3300}) can be written as Eqs.\ (\ref{Einstein 800-3110}). $\Box$

\begin{definition_my}\label{harmonic 800-3400}
If each of the coordinates $x^{\alpha}$ satisfies the following generalized wave equations
\begin{equation}\label{harmonic 800-3410}
\frac{1}{\sqrt{-g_{0}}}\frac{\partial }{\partial x^{\mu}}
\left ( \sqrt{-g_{0}}g^{\mu\nu}\frac{\partial x^{\alpha}}{\partial x^{\nu}}\right )= 0,
\end{equation}
then, we call such a coordinates system harmonic.
\end{definition_my}

In a harmonic coordinates system, we have (\cite{FockV1964}, p.\ 254)
\begin{equation}\label{harmonic 800-3500}
\Gamma^{\nu} = \Gamma^{\mu\nu} = B^{\mu\nu} = B = 0.
\end{equation}

Putting Eqs.\ (\ref{harmonic 800-3500}) into Eqs.\ (\ref{Einstein 800-3110}), we have the following corollary.
\begin{wcorollary_my}\label{field 800-3600}
If we apply the Hilbert gauge Eqs.\ (\ref{Hilbert 500-5800}) and the coordinates system is harmonic, then the field equations Eqs.\ (\ref{Einstein 800-3110}) can be written as
\begin{eqnarray}
G^{\mu\nu}&-&\frac{1}{2g_{0}}\left (\sqrt{-g_{0}}g^{\alpha\beta}-\eta^{\alpha\beta}\right )\frac{\partial^{2}(\sqrt{-g_{0}}g^{\mu\nu})}{\partial x^{\alpha} \partial x^{\beta}} \nonumber\\
&-&\Pi^{\mu,\alpha\beta}\Pi_{\alpha\beta}^{\nu}+\frac{1}{2}y^{\mu}y^{\nu}-\frac{1}{2}g^{\mu\nu}L\nonumber\\
&=&\frac{f_{0}^{2}}{g_{0}}(T^{\mu\nu}-T_{\omega}^{\mu\nu}).\label{Einstein 800-3610}
\end{eqnarray}
\end{wcorollary_my}

We can verify that each of the Galilean coordinates is harmonic. Any constant and any linear function of harmonic coordinates satisfy Eqs.\ (\ref{harmonic 800-3410}). Thus, from Eqs.\ (\ref{Lorentz 700-300}) we see that an inertial coordinate system is harmonic and Eqs.\ (\ref{Einstein 800-3610}) are valid for every inertial system.
In order to study the case of weak fields in inertial systems, we introduce the following assumption.
\begin{assumption_my}\label{assumption 800-3650}
Suppose that the dimensionless parameter $\varpi=m_{0}c/2\rho_{0}q_{0}$ satisfies the following condition
\begin{equation}\label{varpi 800-3660}
\varpi=\frac{m_{0}c}{2\rho_{0}q_{0}}\leq 1.
\end{equation}
\end{assumption_my}

Using the $00$ component of Eqs.\ (\ref{weak 800-3720}) for the case $n=1$ and noticing Eqs.\ (\ref{definition 500-6100}), Eqs.\ (\ref{definition 400-1910}), Eqs.\ (\ref{f0 500-3160}) and Eqs.\ (\ref{definition 200-500}), we have
\begin{equation}\label{velocity 800-3670}
\left | \frac{\partial (2f_{0}\phi^{00})}{\partial (x^{\alpha})} \right |
= \left |\frac{2\rho_{0}q_{0}}{m_{0}c^{2}} \frac{\partial \varphi}{\partial x^{\alpha}} \right |\ll 1.
\end{equation}

Noticing Eq.\ (\ref{Stokes 400-1600}) and using Eq.\ (\ref{velocity 800-3670}) and Eq.\ (\ref{varpi 800-3660}), we have
$| u  | \approx  |\nabla \varphi  |\ll m_{0}c^{2}/(2\rho_{0}q_{0}) \leq c$. Therefore, according to Assumption \ref{assumption 500-4850}, Eqs.\ (\ref{relationship 500-4860}) and Eqs.\ (\ref{relationship 500-4864}) are valid for weak fields.

\begin{wcorollary_my}\label{field 800-3800}
Suppose that (1) the Hilbert gauge Eqs.\ (\ref{Hilbert 500-5800}) is applied on the fields; (2) the filed is weak; (3) Assumption \ref{assumption 500-4850} is valid. Then in an arbitrary inertial coordinate system the field equations Eqs.\ (\ref{Einstein 800-3610}) reduce to
\begin{equation}\label{Einstein 800-3810}
R_{\mu\nu}-\frac{1}{2}g_{\mu\nu}R = \frac{f_{0}^{2}}{g_{0}}T^{m}_{\mu\nu}.
\end{equation}
\end{wcorollary_my}
{\bfseries{Proof of Corollary \ref{field 800-3800}.}}
According to Definition \ref{weak 800-3700}, $f_{0}\phi^{\mu\nu}$ and their first and higher derivatives are small quantities of order $\varepsilon$, where $|\varepsilon| \ll 1$ is a small quantity. Thus, using Eqs.\ (\ref{metric 800-3010}) and Eqs.\ (\ref{expansion 800-3900}), we have the following estimation of the order of magnitude of the following quantities
\begin{equation}\label{order 800-4000}
\sqrt{-g_{0}}g^{\mu\nu}-\eta^{\mu\nu} \sim \varepsilon,
 \quad  \frac{\partial g_{\mu\nu}}{\partial x^{\alpha}}\sim \frac{\partial g^{\mu\nu}}{\partial x^{\alpha}}\sim \varepsilon.
\end{equation}

From Eqs.\ (\ref{metric 800-3010}), we have the following estimation of the order of magnitude of the quantity \begin{equation}\label{order 800-4100}
\frac{\partial^{2}(\sqrt{-g_{0}}g^{\alpha\beta})}{\partial x^{\alpha} \partial x^{\beta}} = \frac{\partial^{2}(-2f_{0}\phi^{\mu\nu})}{\partial x^{\alpha} \partial x^{\beta}} \sim \varepsilon.
\end{equation}

Thus, using Eqs.\ (\ref{order 800-4000}) and Eqs.\ (\ref{order 800-4100}), we have the following estimation of the order of magnitude of the quantity
\begin{equation}\label{order 800-4200}
\left (\sqrt{-g_{0}}g^{\alpha\beta}-\eta^{\alpha\beta}\right )\frac{\partial^{2}(\sqrt{-g_{0}}g^{\alpha\beta})}{\partial x^{\alpha} \partial x^{\beta}} \sim \varepsilon^{2}.
\end{equation}

From Eqs.\ (\ref{Fock 800-2920}) and Eqs.\ (\ref{Fock 800-2930}), we have the following estimation of the order of magnitude of the following quantities
\begin{equation}\label{order 800-4300}
\Pi^{\mu,\alpha\beta} \sim \Pi_{\alpha\beta}^{\nu} \sim \varepsilon.
\end{equation}

Using Eqs.\ (\ref{Fock 800-2960}), we have the following relationship (\cite{FockV1964}, p.\ 143)
\begin{equation}\label{formula 800-4400}
y_{\beta} = \Gamma_{\beta\nu}^{\nu}.
\end{equation}

We also have (\cite{FockV1964}, p.\ 143)
\begin{equation}\label{formula 800-4500}
\Gamma_{\beta\nu}^{\nu} = \frac{1}{2}g^{\mu\nu}\frac{\partial g_{\mu\nu}}{\partial x^{\beta}}.
\end{equation}

From Eqs.\ (\ref{formula 800-4400}), Eqs.\ (\ref{formula 800-4500}) and Eqs.\ (\ref{order 800-4000}),  we have the following estimation of the order of magnitude
\begin{equation}\label{order 800-4600}
y_{\beta} \sim \varepsilon.
\end{equation}

Using Eqs.\ (\ref{Fock 800-2960}) and Eqs.\ (\ref{order 800-4600}),  we have the following estimation of the order of magnitude
\begin{equation}\label{order 800-4700}
y^{\alpha} \sim \varepsilon.
\end{equation}

From Eq.\ (\ref{Fock 800-2970}) and Eqs.\ (\ref{harmonic 800-3500}),  we have
\begin{equation}\label{relation 800-4800}
L = -\frac{1}{2}\Gamma_{\alpha\beta}^{\nu}\frac{\partial g^{\alpha\beta}}{\partial x^{\nu}}.
\end{equation}

Using Eq.\ (\ref{relation 800-4800}), Eqs.\ (\ref{Christoffel 750-1260}) and Eqs.\ (\ref{order 800-4000}),  we have the following estimation of the order of magnitude
\begin{equation}\label{order 800-4900}
L \sim \varepsilon^{2}.
\end{equation}

From Eqs.\ (\ref{order 800-4200}), Eqs.\ (\ref{order 800-4300}), Eqs.\ (\ref{order 800-4700}) and Eq.\ (\ref{order 800-4900}), we see that the second to the fifth term on the right side of Eqs.\ (\ref{Einstein 800-3610}) are all small quantities of order $\varepsilon^{2}$. Ignoring all these small quantities of order $\varepsilon^{2}$ in Eqs.\ (\ref{Einstein 800-3610}) and using Eqs.\ (\ref{relationship 500-4864}), we obtain
\begin{equation}\label{Einstein 800-4950}
G^{\mu\nu} \approx \frac{f_{0}^{2}}{g_{0}}T_{m}^{\mu\nu}.
\end{equation}

Applying the rules of lowering or raising the indexes of tensors, i.e., $G^{\mu\nu}=g^{\mu\lambda}g^{\nu\sigma}G_{\lambda\sigma}$, $T_{m}^{\mu\nu}=g^{\mu\lambda}g^{\nu\sigma}T^{m}_{\lambda\sigma}$, Eqs.\ (\ref{Einstein 800-4950}) can be written as
\begin{equation}\label{Einstein 800-4960}
G_{\lambda\sigma} \approx \frac{f_{0}^{2}}{g_{0}}T^{m}_{\lambda\sigma}.
\end{equation}

Putting Eqs.\ (\ref{Einstein 800-2810}) into Eqs.\ (\ref{Einstein 800-4960}), we obtain Eqs.\ (\ref{Einstein 800-3810}). $\Box$

\begin{wcorollary_my}\label{field 800-5000}
Suppose that the following conditions are valid: (1) the Hilbert gauge Eqs.\ (\ref{Hilbert 500-5800}) is applied on the fields;  (2) the filed is weak; (3) $g_{0}\approx -1$; (4) Assumption \ref{assumption 500-4850} is valid.  Then in an arbitrary inertial coordinate system the field equation Eqs.\ (\ref{Einstein 800-3810}) reduce to
\begin{equation}\label{Einstein 800-5010}
R_{\mu\nu}-\frac{1}{2}g_{\mu\nu}R = -f_{0}^{2} T^{m}_{\mu\nu}.
\end{equation}
\end{wcorollary_my}

If we introduce the following notation
\begin{equation}\label{kappa 800-5100}
\kappa = f_{0}^{2},
\end{equation}
then, Eqs.\ (\ref{Einstein 800-5010}) coincide with Einstein's equations Eqs.\ (\ref{Einstein 100-100}). Thus, we see that the field equations Eqs.\ (\ref{Einstein 800-3055}) are generalizations of the Einstein's equations Eqs.\ (\ref{Einstein 100-100}) in inertial coordinate systems.

\section{Conclusion \label{sec 1100}}
 We extend our previous theory of gravitation based on a sink flow model of particles by methods of special relativistic fluid mechanics. In inertial coordinate systems, we construct a tensorial potential of the $\Omega (0)$ substratum. Based on some assumptions, we show that this tensorial potential satisfies the wave equations. Inspired by the equations of motion of a test particle, a definition of a metric tensor of a Riemannian spacetime is introduced. Generalized Einstein's equations in inertial coordinate systems are derived based on some assumptions. These equations reduce to the Einstein's equations in case of weak field in harmonic coordinate systems. In our theory, gravity is transmitted by the $\Omega (0)$ substratum. The theory of general relativity can not provide a physical definition of the metric tensor of the Riemannian spacetime. In our theory, the background spacetime is the Minkowshi spacetime. However, the flat background spacetime is no longer physically observable. According to the equations of motion of a point particle in gravitational field, to first order, the physically observable spacetime is a Riemannian spacetime. The metric tensor of this Riemannian spacetime is defined based on the tensorial potential of gravitational fields.

\section*{Acknowledgments \label{sec 1200}}
This work was partly supported by the Doctor Research Foundation of Henan Polytechnic University (Grant No. B2012-069). I would like to thank an anonymous reviewer for pointing out some mistakes and providing some suggestions.

\end{document}